\pdfoutput=1 
\documentclass[AMA,STIX1COL]{WileyNJD-v2}

\usepackage{graphicx}
\usepackage{subfigure}

\articletype{RESEARCH ARTICLE}%

\received{}
\revised{}
\accepted{}
        
\begin{document}

\title{A Multi-View Approach Based on Naming Behavioral Modeling for Aligning Chinese User Accounts across Multiple Networks}

\author[1,2]{Junxing Zhu}
\author[1,2]{Xiang Wang*}
\author[2]{Qiang Liu}
\author[1,2]{Xiaoyong Li}
\author[1,2]{Chengcheng Shao}
\author[2]{Bin Zhou}

\authormark{AUTHOR ONE \textsc{et al}}

\address[1]{\orgdiv{College of Meteorology and Oceanography}, \orgname{National University of Defense Technology}, \orgaddress{\state{Hunan}, \country{China}}}
\address[2]{\orgdiv{College of Computer Science and Technology}, \orgname{National University of Defense Technology}, \orgaddress{\state{Hunan}, \country{China}}}

\corres{*Xiang Wang, College of Computer Science and Technology, National University of Defense Technology, Changsha, Hunan, China, 410073. \email{xiangwangcn@nudt.edu.cn}}

\presentaddress{Xiang Wang, College of Computer Science and Technology, National University of Defense Technology, Changsha, Hunan, China, 410073.}

\abstract[Summary]{Hundreds of millions of Chinese people have become social network users in recent years, and aligning the accounts of common Chinese users across multiple social networks is valuable to many inter-network applications, e.g., cross-network recommendation, cross-network link prediction. Many methods have explored the proper ways of utilizing account name information into aligning the common English users' accounts. However, how to properly utilize the account name information when aligning the Chinese user accounts remains to be detailedly studied. In this paper, we firstly discuss the available naming behavioral models as well as the related features for different types of Chinese account name matchings. Secondly, we propose the framework of \textit{\underline{M}ulti-View \underline{C}ross-Network \underline{U}ser \underline{A}lignment (MCUA)} method, which uses a multi-view framework to creatively integrate different models to deal with different types of Chinese account name matchings, and can consider all of the studied features when aligning the Chinese user accounts. Finally, we conduct experiments to prove that \textit{MCUA} can outperform many existing methods on aligning Chinese user accounts between Sina Weibo and Twitter. Besides, we also study the best learning models and the top-$k$ valuable features of different types of name matchings for \textit{MCUA} over our experimental data sets.}


\maketitle

\section{Introduction}
Online social networks are highly developed in recent years \cite{Velayudhan2019Compromised}, and hundreds of millions of Chinese people have become social network users\footnote{Calculated by China Internet Network Information Center(CNNIC), the Statistics Report on China Internet Developing Situation(2015). Published on January, 2015.}. Different social networks may provide different services, so it is natural for individuals to use multiple social networks for different purposes at the same time \cite{Zhu2017cikm}. For example, a Chinese student may use Renren\footnote{www.renren.com} to share funny photos with his classmates, use Sina Weibo\footnote{weibo.com} to follow the latest events, and use Twitter\footnote{www.twitter.com} to connect with international friends. However, the accounts owned by the same user in different social sites are mostly isolated without any correspondence connections to each other \cite{zhang2013predicting}.

Aligning the accounts of common users across different social networks is of great value to many concrete real-world inter-network applications \cite{zhang2013predicting,lu2014identifying,zhang2015integrated,Zhu2017Constrained}. For example, we can recommend new friends or new topics to a new Twitter user according to the social relationship information or personal interest information of his/her related Sina Weibo account, or recommend new products to an Amazon user by analyzing the preferences of his/her close friends in Sina Microblog. The links connecting common users' accounts across different social networks are also referred as ``anchor links''  in some works \cite{zhang2013predicting,Zhu2017Constrained}, and thus the problem of aligning the accounts of common users across multiple networks is also called ``anchor link prediction''.

In recent years, many works have been proposed to align the accounts of common users across different social networks \cite{wang2018user,li2019matching,liu2019structural,wang2019online,li2018matching,ZhangCikm2018,Ma2017Balancing,Zhang2016Social,zhang2015cosnet,liu2014hydra,lu2014identifying,Zafarani2013Connecting,Jain2013,zafarani2009connecting}. And in most of these works \cite{li2019matching,ZhangCikm2018,Ma2017Balancing,Zhang2016Social,zhang2015cosnet,liu2014hydra,lu2014identifying,Zafarani2013Connecting,Jain2013,zafarani2009connecting}, the account name information plays an important role, because:
\begin{itemize}
	\item In general, many people prefer to use similar account names in different social networks since it is not easy for people to remember a large number of different name strings. As one study shows \cite{zafarani2009connecting}, $59\%$ of individuals prefer to use the same account name(s) repeatedly, mostly for ease of remembering.
	\item Unlike the account name information which is sufficient in different networks and is easy to be collected, many other types of information (e.g., the user profile information and the user location information) which can be used to align user accounts are usually very sparse or unavailable in some networks. For example, in some networks due to privacy concerns, many users' profile information is usually anonymized \cite{zhang2015multiple,Xu2014Information,Zhu2017CHRS}. And there are also many users who are not active in some networks, thus they are not likely to post their location information in these networks. As a result, some approaches have to rely on the name information to achieve good performances.
\end{itemize}

Although different ways have been explored to apply the name information matching to the cross-network alignment of user accounts, however, most of them just focus on connecting the accounts of users who mainly use English and create English names (in this paper, these users are referred as English users) \cite{li2019matching,lu2014identifying,Jain2013,zafarani2009connecting,Zafarani2013Connecting,You2011SocialSearch,Backes2018}. Since Chinese users' behavioral models are quite different from English users' when creating the account names, the matching of Chinese name information may encounter some new problems. For example, a Chinese user may use Chinese letters to create his/her account name(s) on Sina Weibo, but use English letters to create his/her Twitter name(s). Although the Twitter username(s) can be the phonetic presentation(s) of the Sina Weibo name(s), the traditional methods which focus on connecting English users can hardly deal with it. 

Recently, a few approaches have been proposed to integrate the matching of name information to the alignment of Chinese user accounts \cite{ZhangCikm2018,zhang2015cosnet,Zhang2016Social,liu2014hydra,Ma2017Balancing,Ma2017Balancing}, but due to the inherent limits of these approaches, there are still two important challenges that remain to be well solved:
\begin{itemize}
	\item \textit{Many new features need to be detailedly studied:} compared with English users, Chinese users usually have different behavioral models when creating the account names. So when matching the Chinese name information, many new features can be extracted to reflect these behavioral models. However, almost all of the existing studies on this topic just focus on a very few Chinese naming behaviors, and thus there are still many behavioral models as well as the related features remain to be detailedly studied.
	\item \textit{Multiple types of matchings need to be dealt with:} since Chinese users often use not only Chinese letters but also English letters to create their account names, their names can be divided into two types. One is \textit{En}, which represents the names without any Chinese letters in them; the other is \textit{Cn}, which represents the names contain Chinese letter(s). As a result, the will be three types of name matchings: both of the matched names belong to \textit{En}, both of the matched names belong to \textit{Cn}, and one name belongs to \textit{En} while the other one belongs to \textit{Cn}. But as far as we know, none of the existing studies have discussed about the differences between these three types of matchings. So how to properly deal with each type of matchings in this way to achieve good performances when aligning Chinese user accounts becomes an important challenge.
\end{itemize}

To solve these challenges, in this paper, we firstly discuss the details of different types of Chinese account name matchings. And then for each type of matchings, we study the available naming behavioral models as well as their related features. Thirdly, we propose the framework of our \textit{\underline{M}ulti-View \underline{C}ross-Network \underline{U}ser \underline{A}lignment (MCUA)} method, which novelly integrates the models of different types of user name matchings into a multi-view framework and can consider all of the studied features. In each time of aligning Chinese user accounts, \textit{MCUA} can use different models to deal with different types of Chinese account name matchings, and then generate a unified result according to the returned results of these models. Finally, we randomly collect the user information of Sina Weibo and Twitter, and then conduct experiments to prove that our \textit{MCUA} method can outperform many existing methods on aligning Chinese user accounts between these two networks. Besides, we also study the best learning models and the top-$k$ valuable features of different matchings for \textit{MCUA} over our experimental data sets.

The remaining part of this paper is organized as follows: At first, we introduce the related works in Section \ref{sec:relate}. The background and preliminaries of our problem are presented in Section \ref{sec:back}. In Section \ref{sec:MUCA}, we introduce our \textit{MCUA} approach. In Section \ref{sec:exp}, we design and conduct the experiments for \textit{MCUA}, and analyze the experimental results. Finally, we conclude in Section \ref{sec:conclusions}.

\section{Related Works}
\label{sec:relate}
In recent years, many works have studied the problem of aligning the user accounts of common users across different networks. Some of them are only based on the user registration information (such as the username, gender, emails) to connect users \cite{li2019matching,Jain2013,Zafarani2013Connecting,liu2013cross,zafarani2009connecting,Iofciu2010Identifying,Backes2018,Motoyama2009I}. While the others try to utilize multiple heterogeneous information for user account alignment, such as user the social relations \cite{wang2018user,liu2019structural,wang2019online,ZhangCikm2018,Ma2017Balancing,Zhu2017Constrained,Zhang2016Social,zhang2015cosnet,kong2013inferring}, user interests \cite{wang2018user,wang2019online,Nie2016Identifying,lu2014identifying}, user temporal distribution features \cite{li2018matching,lu2014identifying,Zhu2017Constrained,kong2013inferring,Lu2018Releasing}.

To most of the studies on user account alignment \cite{li2019matching,ZhangCikm2018,Ma2017Balancing,Zhang2016Social,zhang2015cosnet,liu2014hydra,lu2014identifying,Zafarani2013Connecting,liu2013cross,zafarani2009connecting,Iofciu2010Identifying,Backes2018,Motoyama2009I}, the account name information is very important, since many users like to assign their accounts in different networks with very similar names, and the account names in most networks are very easy to be acquired. And how to properly utilize the name information in the alignment of accounts owned by the English users have already been well studied by many works \cite{li2019matching,liu2014hydra,Zafarani2013Connecting,zafarani2009connecting,Iofciu2010Identifying,Backes2018,Motoyama2009I}, among them: Vosecky J. et al. \cite{Vosecky2009User} propose a method which based on web profile matching to connect users between Facebook and StudiVZ. In their study, they compare 3 kinds of name matching algorithms, and select the best one for profile matching. However, the names used to test the performances of these algorithms are manually generated and owned by only one person. And in real world applications, the situation can be much more complicated, and thus a large number of name samples are needed to conduct the comparison.
Zafarani and Huan Liu \cite{zafarani2009connecting} first introduce a methodology for connecting user accounts across social networks by usernames. And then in 2013, they observed that humans tend to have consistent behavior patterns when naming their account in different social networks \cite{Zafarani2013Connecting}, so they conduct a very detailed study on these behavior patterns as well as the related naming behavioral features that can be used to connect user accounts across social networks. However, their later work mainly bases on the assumption that multiple prior usernames of the same individuals are available, but it is not easy to acquire the prior names for a given user account in most cases. C. Lu et al. \cite{lu2014identifying} propose a methodology which utilizes different information to connect user accounts across different networks for potential marketing applications. In their study, five features that can be calculated by the single prior name are selected to be used in their method. Tereza Iofciu et al. use username and user tags to match user accounts across Flickr, Delicious and Stumble-Upon \cite{Iofciu2010Identifying}, by comparing five username similarity metrics' performances on matching user accounts across these three sites, they pick out the most suitable username similarity metric for their application at last. Y. Li et al. \cite{li2019matching} propose a model UISN-UD and a two-stage implementation framework for matching user accounts across social networks. Their proposed model is based on username and display name, and uses the longest common substring to evaluate the name similarity.

Since Chinese users' behavioral models are quite different from English users' when creating the account names, matching Chinese account names can be very different from matching English account names. Although there are several works trying to apply the account name information to the alignment of Chinese user accounts in recent years \cite{ZhangCikm2018,liu2014hydra,Zhang2016Social,liu2013cross,Ma2017Balancing,zhang2015cosnet}, it is a pity that most of them just focus on a very few Chinese naming behaviors, which may be not sufficient enough to cover most of the common situations of matching Chinese user account names. Among them, S. Liu et al. \cite{liu2014hydra} propose a framework which can connect user accounts across heterogeneous Chinese social media platforms by using multiple user features, and then use data from five Chinese social media platforms to demonstrate that their framework can perform very well in Chinese user account alignment. However, they just directly conduct the username matching without any analysis of the multiple naming behavior models of different users. Y. Zhang et al. \cite{Zhang2016Social} develop a method that can align user accounts across Chinese social networks, and to better utilize the user nickname information, they evaluate the relevance of the nicknames owned by the same users and novelly create a new feature which can be used to deal with three common cases of Chinese nickname matching very well. However, in real scenarios, the common cases of Chinese account name matching are not just only three. D. Liu et al.\cite{liu2013cross} design a methodology to find the corresponding username(s) for a specific Chinese username. They compare three name similarity computation methods on matching Chinese user names, and select the best one for their method. However, the names used to test the performances of these three methods are manually generated potential usernames of only one person. 
Several state-of-art methods \cite{zhang2015cosnet,Ma2017Balancing,ZhangCikm2018} explore the ways of utilizing user social information and user text information into the alignment of user accounts, and can perform well on some Chinese social networks. However, they just directly proposed their username matching methods based on models like TF-IDF\cite{zhang2015cosnet,Ma2017Balancing}, CNN\cite{ZhangCikm2018}, etc. without detailedly studying the multiple naming behavior models of Chinese users. 

Multi-view learning has been widely studied in recent years \cite{Cao2015Tensor}. The multi-view here can refer to the various descriptions of a given sample, and thus the goal of multi-view learning is to properly fuse the different descriptions in the learning process \cite{Luo2015Multiview}. Many studies have applied multi-view learning to different applications, such as classification \cite{Zhu2016Block}, retrieval \cite{Liu2015Multiview}, clustering \cite{Nie2016Parameter}, etc. According to the level of the fusion being carried out, the multi-view classification methods can be grouped into two major categories \cite{Luo2015Multiview}: feature level
fusion based methods and classifier-level fusion based methods. The feature level fusion based methods often directly fuses different types of features together, e.g., concatenate the different kinds of features into a long vector, and then use the learning model to process the fused information \cite{Mcfee2011Learning,Xu2015Multi}. While the classifier-level fusion based methods assign different views with their own classifiers and conduct the fusion process at the classifier-level. For example, fusing the outputs or decisions of different views' classifiers \cite{Wozniak2009Some}, or communicate information with other views when learning classifier of the current view \cite{Sindhwani2005A}. And since several researches have proved that classifier-level fusion outperforms simple feature concatenation in multi-view classification area \cite{Luo2015Multiview,Kludas2008Information}, we apply the classifier-level fusion based multi-view learning method to the area of user account alignment in this paper. 

\section{Background and Preliminaries}
\label{sec:back}
In this section, we illustrate the background and preliminaries of this study. However, before the illustration, we summarize the main notations used in this paper in Table \ref{tab:main_notation}.
\begin{table}[htb]
	\centering
	\caption{Main notations.}
	\label{tab:main_notation}       
	\begin{tabular}{p{2.5cm}p{5.5cm}}
		\hline\noalign{\smallskip}
		Notation& Description\\	
		\noalign{\smallskip}\hline\noalign{\smallskip}
		$\mathcal{G}^{(a)}$& The $a$th information networks\\
		$\mathcal{U}^{(a)}$& The user account set of $\mathcal{G}^{(a)}$\\
		$u^{(a)}_i$, $u^{(a)}_j$& The $i$th and $j$th user account in $\mathcal{U}^{(a)}$\\
		$N^{(a)}_i$ & The set of names used by the $u^{(a)}_i$\\
		$n^{(a)}_{i,k}$ & The $k$th name in $N^{(a)}_i$\\
		$M$ & The numbers of user accounts in $\mathcal{G}^{(1)}$\\
		$N$ & The numbers of user accounts in $\mathcal{G}^{(2)}$\\
		$\mathcal{A}=\{A_{i,j}\}^{M\times N}$ & The set of alignment relationships between the user accounts in $\mathcal{G}^{(1)}$ and $\mathcal{G}^{(2)}$\\
		$A_{i,j}$ & The alignment relationships between the user accounts $u^{(1)}_i$ and $u^{(2)}_j$\\
		\noalign{\smallskip}\hline
	\end{tabular}
\end{table}

Suppose there are two network $\mathcal{G}^{(1)}$ and $\mathcal{G}^{(2)}$, the user account sets of $\mathcal{G}^{(1)}$ and $\mathcal{G}^{(2)}$ are $\mathcal{U}^{(1)}= \{u^{(1)}_1,u_2^{(1)},\cdots,u_M^{(1)}\}$ and $\mathcal{U}^{(2)}=\{u^{(2)}_1,u_2^{(2)},\cdots,u_N^{(2)}\}$ respectively. Where $u^{(1)}_i$ and 
$u^{(2)}_j$ are the $i$th user account in $\mathcal{G}^{(1)}$ and the $j$th user account in $\mathcal{G}^{(2)}$ respectively. $M$ and $N$ denote the numbers of user accounts in $\mathcal{G}^{(1)}$ and $\mathcal{G}^{(2)}$. $N^{(a)}_i$ denotes the set of names used by the $u^{(a)}_i$, and $n^{(a)}_{i,k}$ is the $k$th name in $N^{(a)}_i$, where $a \in \{1,2\}$. And $\mathcal{A}=\{A_{i,j}\}^{M\times N}$ denotes the set of alignment relationships between the user accounts in $\mathcal{G}^{(1)}$ and $\mathcal{G}^{(2)}$, if $u^{(1)}_i$ and $u^{(2)}_j$ are (or are not) owned by the same user, then the value of $A_{i,j}$ is set as $1$ (or $0$), and we label it as ``positive'' (``negative''); and if we don't know whether $u^{(1)}_i$ and $u^{(2)}_j$ belong to the same user or not, the value of $A_{i,j}$ is unknown, and thus it is an unlabeled alignment relationship. The task of user alignment across $\mathcal{G}^{(1)}$ and $\mathcal{G}^{(2)}$ is to create a model and use it to predict the value of each unknown $A_{i,j}$ in $\mathcal{A}$. Specifically, the goal of our research is to propose a user alignment model, which can properly use the information extracted from the names of Chinese user accounts and achieve good performances on aligning these accounts.

\section{Multi-View Approach for Aligning Chinese User Accounts across Multiple Networks}
\label{sec:MUCA}
In this section, we introduce the \textit{\underline{M}ulti-View \underline{C}ross-Network \underline{U}ser \underline{A}lignment (MCUA)} method, which aims at aligning Chinese user accounts across multiple networks based on naming behavioral modeling. Firstly, we analyze different types of Chinese name matchings, with which we will be confronted when aligning Chinese user accounts across multiple networks. Then to each type of Chinese name matchings, we discuss the available features that are suitable to be used to train its related name matching model. Finally, we propose the framework of \textit{MCUA}, which fuses the results of all the trained Chinese name matching models, in this way to better solve the alignment of Chinese user accounts. The details are shown in the following subsections.
\subsection{Multiple types of Chinese Name Matchings}
\label{subsec:mtcnm}
As mentioned before, different from English users, a Chinese user may use Chinese letters to create one account name, but use English letters to create his/her other account names, thus we can divide the Chinese account names into the following two types:
\begin{itemize}
	\item \textit{En}: the names created by Chinese users but contain no Chinese letters. These names include the English names of Chinese users, the phonetic presentation of the users' Chinese names, and the abbreviates formed by the first letters of the phonetic presentations of all the Chinese letters in users' Chinese names, etc.. 
	\item \textit{Cn}: the account names contain one or more Chinese letters. There are two forms of these names, one is the names only formed by Chinese letters, such as a user's Chinese name. The other form is the names not only contain Chinese letters but also contain other letters, which can be some English prefixes like ``Mr.'', ``Prof.'' and ``Dr.'', or some non-Chinese letters which have no obvious meanings in them (e.g., some randomly selected numbers, and some special symbols like ``*'', ``\$'' and ``\#'', etc.).
\end{itemize}

Thus there are three types of Chinese account name matchings: 
\begin{itemize}
	\item \textit{EE}: all of the account names to be matched by it are \textit{En} names.
	\item \textit{CC}: all of the account names to be matched by it are \textit{Cn} names.
	\item \textit{CE}: for any two account names to be matched by it, one is \textit{Cn}, and the other one is \textit{En}.
\end{itemize}

We notice that a few existing approaches have studied the matching of English user account names \cite{zafarani2009connecting,Zafarani2013Connecting,Iofciu2010Identifying}, which looks very similar to the \textit{EE}. However, since in the \textit{EE} matching, all of the account names are created by Chinese users, there must be some special Chinese user behavioral patterns which make the \textit{EE} matching different from the traditional English account name matchings. For example, English users like using word-splitting symbols like blank spaces or ``\_'' to split each part of their names, thus these symbols are very important to the matching of English account names. However, Chinese users need not to split each part of their names by these symbols, thus they may casually use blank spaces or ``\_'' when creating the \textit{En} names, so deleting all the word-splitting symbols in each name may help us extract better features that will be used in the \textit{EE} matching. 

\subsection{The available features for each type of Chinese name matchings}
As far as we know, many works have detailedly studied the features for account name matchings \cite{zafarani2009connecting,Zafarani2013Connecting,Iofciu2010Identifying,liu2013cross}. Among them, Zafarani et al. \cite{Zafarani2013Connecting} have detailedly studied the features that can be used to match English usernames, however, many of these studied features should be extracted from the prior usernames of each identity, but it is not easy to get one person's prior usernames in most circumstances, so their method can hardly be applied to many other applications. Liu et al. \cite{liu2013cross} studies the features for matching Chinese usernames, however, it's approach is an unsupervised approach without discussing different types of Chinese account name matchings. Besides, the performances of its studied username similarity algorithms are just tested by the manually generated potential usernames of only one Chinese user. Different from these approaches, for each type of Chinese name matchings in subsection \ref{subsec:mtcnm}, we discuss several available features that can be used to train its matching model.
\subsubsection{The available features that can be used in \textit{EE}}
Among the three types of Chinese account name matchings, since \textit{EE} matches the names which contain no Chinese letter, we need not study the features about the Chinese letters contained in the account names. Therefore, we discuss the features that can be used in \textit{EE} as follows:

\textit{Account name similarity}: According to the previous researches \cite{zafarani2009connecting,Zafarani2013Connecting}, due to the limits of time, memory and knowledge, humans are likely to create the same or very similar account names. Thus the account name similarity is a very important feature for account name matchings. And different from most English users, many Chinese users casually use upper case and lower case English letters when creating their account names. Therefore, here we study two types of account name similarities:
\begin{enumerate}[1.]
	\item The similarity computed from two names directly.
	\item The similarity computed from two names, whose upper case letters have been transformed to the lower case letters firstly.
\end{enumerate}

The account name similarity is computed from the Levenshtein distance, which denotes the minimum number of single-letter edits (insertions, deletions or substitutions) required to change one name into the other name. For two account names $n^{(1)}_{a,b}$ and $n^{(2)}_{c,d}$, if their Levenshtein distance is $LD(n^{(1)}_{a,b},n^{(2)}_{c,d})$, then their similarity is computed as:
\begin{equation}
\label{eq:ld_sim}
\begin{aligned}
sl(n^{(1)}_{a,b},n^{(2)}_{c,d})=1-\frac{LD(n^{(1)}_{a,b},n^{(2)}_{c,d})}{\max(|n^{(1)}_{a,b}|,|n^{(2)}_{c,d}|)}
\end{aligned}
\end{equation}
where $|n^{(k)}_{i,j}|$ is the length of account name $n^{(k)}_{i,j}$.

\textit{The proportion of the longest common substring}:
The account names of a given user usually share the same substring, it can be: 1) the string of user's personal information (name, company, gender, and role, etc.) 2)the string of something meaningful to the user (e.g., a female who loves Disney may select disney as the necessary parts of her account names). And for two given account names, if their \textit{longest common substring} takes a large proportion in each of them, they are likely to belong to the same user. Therefore we can use the proportion of the \textit{longest common substring} of two given account names to analyze whether the two names belong to the same user. Since the extraction of the \textit{longest common substring} is very sensitive to the change of letters, we can extract four types of the \textit{longest common substring} from any two given account names:
\begin{enumerate}[1.]
	\item The \textit{longest common substring} extracted directly.
	\item The \textit{longest common substring} extracted after all the word-splitting symbols have been deleted in both of these two names.
	\item The \textit{longest common substring} extracted after all the upper case letters have been transformed to the lower case letters in both of these two names.
	\item The \textit{longest common substring} extracted after all the word-splitting symbols have been deleted and all the upper case letters have been transformed to the lower case letters in both of these two names.
\end{enumerate}
Suppose that the \textit{longest common substring} of two account names $n^{(1)}_{a,b}$ and $n^{(2)}_{c,d}$ is $LCS(n^{(1)}_{a,b},n^{(2)}_{c,d})$, the proportion of the \textit{longest common substring} of both $n^{(1)}_{a,b}$ and $n^{(2)}_{c,d}$ is:
\begin{equation}
\label{eq:pls_sim}
\begin{aligned}
pls(n^{(1)}_{a,b},n^{(2)}_{c,d})=\frac{2* LCS(n^{(1)}_{a,b},n^{(2)}_{c,d})}{|n^{(1)}_{a,b}|+|n^{(2)}_{c,d}|}
\end{aligned}
\end{equation}
Since there are four types of \textit{longest common substring}, we can use Eq. (\ref{eq:pls_sim}) to compute four types of the proportion of the \textit{longest common substring} for any two account names.

\textit{The similarity of special symbols}: Some users like to use some special symbols which are not Chinese or English letters in their account names, these symbols can be Arabic numerals, Acrophonic numerals and punctuation, etc. For example, a man who was born in 1988 may name himself \textit{Jack1988}, and a user who want to be rich may select \textit{Show\_Me\_\$\$\$} to be the account name. So analyzing the common special symbols of two account names can help us predict whether these two names belong to the same user. If we extract the special symbols in account name $n^{(1)}_{a,b}$, and use them to form the string $sp(n^{(1)}_{a,b})$ according to their orders in $n^{(1)}_{a,b}$. And then form $sp(n^{(2)}_{c,d})$ for account name $n^{(2)}_{c,d}$ in a similar way. We can analyze two features from the common special symbols in $n^{(1)}_{a,b}$ and $n^{(2)}_{c,d}$ as follows:
\begin{enumerate}[1.]
	\item The \textit{cosine similarity} of $sp(n^{(1)}_{a,b})$ and $sp(n^{(2)}_{c,d})$, which represents the similarity of the distribution of special symbols in $n^{(1)}_{a,b}$ and $n^{(2)}_{c,d}$.
	\item The \textit{Jaccard index} of $sp(n^{(1)}_{a,b})$ and $sp(n^{(2)}_{c,d})$, which denotes the percentage of the common special symbols in all the special symbols of $n^{(1)}_{a,b}$ and $n^{(2)}_{c,d}$.
\end{enumerate}

\textit{The similarity of abbreviations}: Some users may select some abbreviations to create their account names, e.g., the abbreviations of their own names, their company names, and their occupations. For a Chinese user who is named \textit{Lei Li} (the phonetic presentation), and works for \textit{Microsoft}, his account names can be \textit{LeiLi@Microsoft}, or some abbreviated forms like \textit{LL@MS}, \textit{Li@MS}, etc. Although a Chinese user can use different forms of abbreviations to create his/her account names in different social networks, however, by computing the \textit{longest common subsequence} of these names, we can still discover the similarities among them. Here we can extract two types of the \textit{longest common subsequence} from any two given account names:
\begin{enumerate}[1.]
	\item The \textit{longest common sequence} extracted directly.
	\item The \textit{longest common sequence} extracted after all the upper case letters have been transformed to the lower case letters in both of these two names.
\end{enumerate}
Let $LCQ(n^{(1)}_{a,b},n^{(2)}_{c,d})$ denote the extracted \textit{longest common subsequence} of account names $n^{(1)}_{a,b}$ and $n^{(2)}_{c,d}$ which may be in their abbreviated forms, we can compute the similarity of their abbreviation(s) as follows:
\begin{equation}
\label{eq:lcq_sim}
\begin{aligned}
sa(n^{(1)}_{a,b},n^{(2)}_{c,d})=\frac{2\times LCQ(n^{(1)}_{a,b},n^{(2)}_{c,d})}{|n^{(1)}_{a,b}|+|n^{(2)}_{c,d}|}
\end{aligned}
\end{equation}
Since there are two types of the \textit{longest common subsequence} for \textit{EE}, we can get two types of abbreviation similarities.

\textit{The similarity of the non-special letters}:
For most of the Chinese user account names, their main parts are formed by the Chinese letters and English letters, which are noted as the non-special letters in this paper (for the \textit{En} names, their non-special letters are only the English letters). So when matching two Chinese user account names, it is important to analyze the similarity of the non-special letters contained in each of them. And for \textit{EE}, we should analyze the values of extracting similarity features from the English letters in \textit{En} names. Let $ns(n_a)$ denote the string formed by all of a given account name $n_a$'s non-special letters according to their orders in $n_a$ (e.g., if $n_a=$``12Jack\_Wu'', then $ns(n_a)=$``JackWu''). So for two given names $n^{(1)}_{a,b}$ and $n^{(2)}_{c,d}$, we can extract the features of the similarity of $ns(n^{(1)}_{a,b})$ and $ns(n^{(2)}_{c,d})$ as follows:
\begin{enumerate}[1.]
	\item The similarity of the non-special letter distribution which is computed from the cosine similarity of $ns(n^{(1)}_{a,b})$ and $ns(n^{(2)}_{c,d})$.
	\item The percentage of the common non-special letters in all the non-special letters of $n^{(1)}_{a,b}$ and $n^{(2)}_{c,d}$, which is computed from the \textit{Jaccard index} of $ns(n^{(1)}_{a,b})$ and $ns(n^{(2)}_{c,d})$.
	\item The proportion of the \textit{longest common substring} in both $ns(n^{(1)}_{a,b})$ and $ns(n^{(2)}_{c,d})$, which is computed by $pls(ns(n^{(1)}_{a,b}),ns(n^{(2)}_{c,d}))$ according to Eq. (\ref{eq:pls_sim}).
	\item The similarity of $ns(n^{(1)}_{a,b})$ and $ns(n^{(2)}_{c,d})$, which is computed from $sl(ns(n^{(1)}_{a,b}),ns(n^{(2)}_{c,d}))$ according to Eq. (\ref{eq:ld_sim})
\end{enumerate}
Since we can not only directly extract features from the letters in $ns(n^{(1)}_{a,b})$ and $ns(n^{(2)}_{c,d})$, but also extract these features after transforming all the upper case English letters to the lower case English letters, in total, there are $4 \times 2$ features to be extracted here.

\subsubsection{The available features that can be used in \textit{CE}}
As we discussed in \ref{subsec:mtcnm}, for any two account names matched by \textit{CE}, one is \textit{Cn}, and the other is \textit{En}. For a lot of \textit{Cn} names, the non-Chinese letters can still occupy a large proportion in the letters that make up them. So the available features for \textit{EE} can also be use in \textit{CE}. However, as the Chinese letters also take an important position in most \textit{Cn} names, but these Chinese letters in \textit{Cn} names cannot be directly used to match the non-Chinese letters in \textit{En} names, we should explore how to discover the possible \textit{En} forms (phonetic presentations) of the given \textit{Cn} names, and then extract some valuable similarity features for \textit{CE} by comparing these \textit{En} forms with the given \textit{En} names.

\begin{figure}
	\centering
	\includegraphics[width=8cm,height=4.0cm]{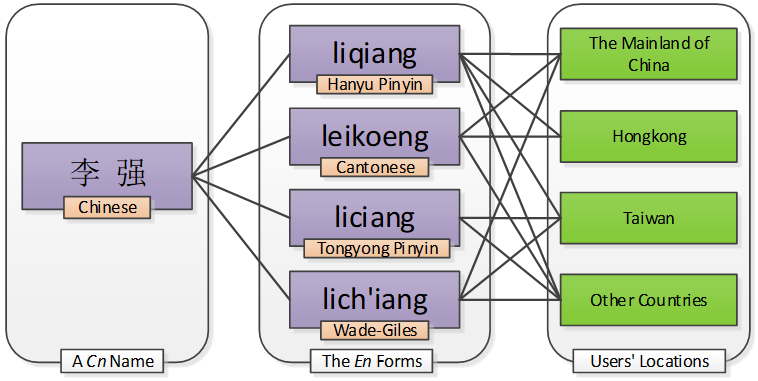}
	\caption{The example of transforming a \textit{Cn} name to multiple \textit{En} names by different romanization systems, and the locations in which a romanization system may be used by some users.}
	\label{fig:multi-phonetic}
\end{figure}

\begin{figure*}
	\centering
	\includegraphics[width=10cm,height=3.5cm]{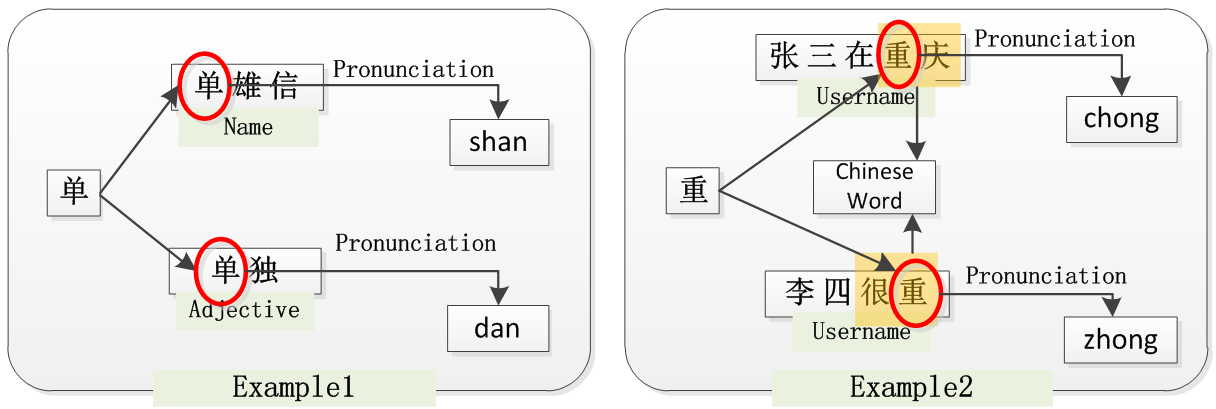}
	\caption{Two examples of polyphone letters, which have multiple pronunciations and can be represented by different phonetic combinations in different contexts. Here we use \textit{Hanyu Pinyin} to illustrate these phonetic combinations.}
	\label{fig:polyphone}
\end{figure*}

When creating the account names, there are several naming factors that make different Chinese users use different strings of English letters to represent a specific Chinese letter. Therefore, before we discovering the possible \textit{En} forms of the given \textit{Cn} names, we should firstly discuss these factors of creating Chinese account names. Here we focus on three important naming factors and briefly discuss the corresponding latent relationships as follows:
\begin{itemize}
	\item \textit{Multiple Phonetic Transcriptions}: Chinese users from the mainland of China, Hongkong, Taiwan and other countries use different romanization systems. In this paper, we focus on the four most popular romanization systems in China, e.g., \textit{Hanyu Pinyin}, \textit{Cantonese}, \textit{Tonyong Pinyin} and \textit{Wade-Giles}. Figure \ref{fig:multi-phonetic} shows an example of transforming a given $Cn$ name to four $En$ names according to the four romanization systems, and the users' main locations of each romanization system.
	\item \textit{Many Polyphone Letters}: Many Chinese letters are polyphone letters. As the two examples illustrated in Figure \ref{fig:polyphone}, each polyphone letter has multiple pronunciations, and can be represented by different phonetic combinations in different contexts. So in order to translate the Chinese polyphone letters to the right phonetic combinations when inferring the possible \textit{En} forms of the given \textit{Cn} names, we should analyze the contexts of these polyphone letters first.
	\item \textit{Different Orders of Family Names}: According to the traditional rules of Chinese, family names should be written before given names. However, many Chinese prefer to use their Chinese names' phonetic presentations to be their English names, and in the phonetic presentations, family names may not only be written before, but also behind the given names. So there exist two kinds of family name orders when transforming the \textit{Cn} names to the \textit{En} names.
\end{itemize}
So in total, after we ensure the right pronunciations of the Chinese polyphone letters contained in the \textit{Cn} names according to their contexts, since the orders of family names are independent to the romanization systems, there are $4 \times 2$ kinds possible \textit{En} forms can be discovered for the \textit{Cn} names. Supposing that $Hy(n)$, $Ct(n)$, $Wd(n)$, and $Ty(n)$ denote the \textit{Hanyu Pinyin} form, \textit{Cantonese} form, \textit{Tonyong Pinyin} form and \textit{Wade-Giles} form of a Chinese $n$ respectively, for each \textit{Cn} name $n_c$, we can generate its possible \textit{En} forms as follows:
\begin{enumerate}[Step 1:]
	\item Create a set $S_F$ which consists of almost all the Chinese family names. 
	\item Since the family names can only exist at the beginning of the names formed by Chinese letters, we examine whether the beginning of $n_c$ is a family name which is contained in $S_F$. If it is, we can move the discovered family name from the beginning to the end of $n_c$, in this way to form a name $\dot{n}_c$.
	\item Create a hash table $H_c$ which can map almost all the Chinese non-polyphone letters to their phonetic forms of the four romanization systems in Figure \ref{fig:multi-phonetic}.
	\item Create a transition function $trp(n)$ which can properly transform almost all the polyphone letters that may exist in Chinese names to their related phonetic forms of the four romanization systems in Figure \ref{fig:multi-phonetic} according to their contexts.
	\item Use $trp(n_c)$ to transform all the polyphone letters in $n_c$ to their phonetic forms and then use $H_c$ to transform the rest Chinese letters to their phonetic forms. Since there exist four romanization systems, as it shows in Figure \ref{fig:multi-phonetic}, we can get four transformed names for $n_c$, which are $Hy(n_c)$, $Ct(n_c)$, $Wd(n_c)$, and $Ty(n_c)$.
	\item If $\dot{n}_c$ exists, transform it to $Hy(\dot{n}_c)$, $Ct(\dot{n}_c)$, $Wd(\dot{n}_c)$, and $Ty(\dot{n}_c)$ in the way similar to Step $5$.
\end{enumerate}
By conducting this preprocess, we can get eight phonetic presentations of $n_c$ (i.e., $Hy(n_c)$, $Ct(n_c)$, $Wd(n_c)$, $Ty(n_c)$, $Hy(\dot{n}_c)$, $Ct(\dot{n}_c)$, $Wd(\dot{n}_c)$, and $Ty(\dot{n}_c)$). In addition, as shown in Figure \ref{fig:polyphone}, there exist two types of Chinese polyphone letters. One type is formed by the special phonetic letters whose pronunciations in the family names are different from their pronunciations in other contexts (See Example $1$), and the other type is the normal phonetic letters which have multiple pronunciations in different words (See Example $2$). So the transform function $trp(n)$ has two hash table $H_f$ and $H_w$, $H_f$ maps almost all the polyphone letters that can be used as Chinese family names to their related phonetic forms of the four romanization systems. $H_w$ maps almost all the Chinese words which contain polyphone letter(s) to their related phonetic forms of the four romanization systems. And in the transform process, $trp(n)$ will firstly use $H_f$ to discover the polyphone letters in the family name of $n_c$ (or $\dot{n}_c$), then use $H_w$ to discover the Chinese words that contain polyphone letter(s) in the rest part of $n_c$ (or $\dot{n}_c$), finally transform all the examined polyphone letters or words to their right phonetic forms according to $H_f$ and $H_w$. 

After getting the possible phonetic presentations for the \textit{Cn} names, we can extract the same new similarity features for \textit{CE}. Suppose $n^{(1)}_{a,b}$ is a \textit{Cn} name in network $G^{(1)}$ and $n^{(2)}_{c,d}$ is an \textit{En} name in network $G^{(2)}$, while $Pf(n^{(1)}_{a,b})$ is one of the eight possible phonetic forms of $n^{(1)}_{a,b}$. Then we can extract these new features from $Pf(n^{(1)}_{a,b})$ and $n^{(2)}_{c,d}$. And since the upper case and lower case English letters can be casually used when transforming $n^{(1)}_{a,b}$ to $Pf(n^{(1)}_{a,b})$, we set all the upper case letters in $Pf(n^{(1)}_{a,b})$ and $n^{(2)}_{c,d}$ to the related lower case letters before extracting the features. The features to be extracted are as follows:

\begin{enumerate}[1.]
	\item \textit{The Cosine Similarity between the En name and the transformed Cn name:} Similar to extract the account name similarity for \textit{EE}, we use Eq. (\ref{eq:ld_sim}) to compute $sl(Pf(n^{(1)}_{a,b}),n^{(2)}_{c,d})$, which is the similarity between $Pf(n^{(1)}_{a,b})$ and $n^{(2)}_{c,d}$. 
	\item \textit{The proportion of the longest common substring in the En name and the transformed Cn name:} By using the Eq. (\ref{eq:pls_sim}), we can compute the proportion of the \textit{longest common substring} for $Pf(n^{(1)}_{a,b})$ and $n^{(2)}_{c,d}$. Similar to extract the proportion of the longest common substring the proportion of the longest common substring for \textit{EE}, considering whether to eliminate the influence of word-splitting symbols in the names or not, we can extract two features here.
	\item \textit{The similarity of the abbreviations in the En name and the transformed Cn name:} By using the Eq. (\ref{eq:lcq_sim}), we can compute the similarity of the abbreviations from  $Pf(n^{(1)}_{a,b})$ and $n^{(2)}_{c,d}$, which is $sa(Pf(n^{(1)}_{a,b}),n^{(2)}_{c,d})$.
	\item \textit{The similarity of the non-special letters in the En name and the transformed Cn name:} Here we firstly extract the strings of English letters $el(Pf(n^{(1)}_{a,b}))$ and $el(n^{(2)}_{c,d})$ from $Pf(n^{(1)}_{a,b})$ and $n^{(2)}_{c,d}$. And then by using the same ways of extracting the four features of \textit{the similarity of the non-special letters} for \textit{EE}, we can get four features from $el(Pf(n^{(1)}_{a,b}))$ and $el(n^{(2)}_{c,d})$.  
\end{enumerate}
Where the four features of \textit{the similarity of the non-special letters} are: 1) the similarity of the English letter distributions in $Pf(n^{(1)}_{a,b})$ and $n^{(2)}_{c,d}$, which is computed from the cosine similarity of $el(Pf(n^{(1)}_{a,b}))$ and $el(n^{(2)}_{c,d})$; 2) the percentage of common English letters in all the English letters of $Pf(n^{(1)}_{a,b})$ and $n^{(2)}_{c,d}$, which is computed from the \textit{Jaccard index} of $el(Pf(n^{(1)}_{a,b}))$ and $el(n^{(2)}_{c,d})$; 3) the proportion of the \textit{longest common substring} in both $el(Pf(n^{(1)}_{a,b}))$ and $el(n^{(2)}_{c,d})$, which is computed by $pls(el(Pf(n^{(1)}_{a,b})),el(n^{(2)}_{c,d}))$ according to Eq. (\ref{eq:pls_sim}); and 4) the similarity of $el(Pf(n^{(1)}_{a,b}))$ and $el(n^{(2)}_{c,d})$, which is computed from $sl(el(Pf(n^{(1)}_{a,b})),el(n^{(2)}_{c,d}))$ according to Eq. (\ref{eq:ld_sim}).

For each kind of the possible phonetic forms of \textit{Cn} names, we can use them to extract the above eight features for \textit{CE}, and there are eight possible phonetic forms of \textit{Cn} names; so we can get $8 \times 8$ features by considering the possible phonetic forms of \textit{Cn} names. And as we discussed at the beginning of this subsection, since the $18$ features for \textit{EE} may adapt to \textit{CE}, in total there are $64+18$ available features for \textit{CE}.

\subsubsection{The available features that can be used in \textit{CC}}
As we discussed in subsection \ref{subsec:mtcnm}, both of the two account names matched by \textit{CC} are \textit{Cn}. And similar to \textit{CE}, the $18$ available features for \textit{EE} can be directly used in \textit{CC}. However, since the Chinese letters also have an important position to most \textit{Cn} names, apart from directly extracting features from these Chinese letters like what we do for \textit{EE}, how to fully consider the properties of Chinese letters in \textit{Cn} names, and properly extract the features about the Chinese letters for \textit{CC} should also be explored. And there are several cases of the \textit{Cn} names, which are owned by the same user but cannot be effectively aligned by just using the $18$ available features for \textit{EE}. We list these cases as follows:
\begin{enumerate}[\text{Case} 1:]
	\item 
	Since Chinese letters can be presented in different forms (e.g., the simplified Chinese letters, and the traditional Chinese letters), some Chinese users use the simplified Chinese letters to form some of their account names, but use the traditional Chinese letters to form their other account names.
	\item There are some Chinese users, each of which uses some Chinese letters to create his/her account name $n^{(1)}_c$, and uses the other Chinese letters which are homophonic to the letters in $n^{(1)}_c$ to create his/her other names.
	\item $n^{(1)}_c$ and $n^{(2)}_c$ are two \textit{Cn} names in different networks that are owned by the same user. $n^{(1)}_c$ contains some Chinese letters which are represented by their phonetic forms in $n^{(2)}_c$.
\end{enumerate}

To deal with Case 1, we can transform all the traditional Chinese letters in \textit{Cn} names to the simplified Chinese letters before extracting the features for name alignment. With the help of Chinese dictionary, we can firstly create a table $T_c$, which maps almost all the traditional Chinese letters that may exist in \textit{Cn} names to the simplified Chinese letters, and then use $T_c$ to conduct the transform. Suppose $n_c$ is a \textit{Cn} name which contains some traditional Chinese letters, after using $T_c$ to transform all the traditional letters in $n_c$ to the simplified Chinese letters, we can get its transformed form $Ts(n_c)$. For Case 2 and Case 3, we can transform all the Chinese letters in \textit{Cn} names to their phonetic presentations before extracting the features. Similar to the way of transforming the \textit{Cn} names to their \textit{En} forms in the process of extracting features for \textit{CE}, we can transform a given \textit{Cn} name $n_c$ to its four \textit{En} forms $Hy(n_c)$, $Ct(n_c)$, $Wd(n_c)$, and $Ty(n_c)$ according to the four romanization systems.

In addition, we notice all the account names processed by \textit{CC} are the \textit{Cn} names, in each of which the family name (if it exists) is written before the given name according to the Chinese naming behaviors. And thus for \textit{CC}, we will not consider the order of family names (i.e., consider transforming a given \textit{Cn} name $n_c$ to $\dot{n}_c$, whose family name is written behind the given name). Suppose $t(n_c)$ is a given function which can transform the \textit{Cn} name $n_c$ to a specific name form (i.e., $t(n_c)$ can be $Ts(n_c)$, $Hy(n_c)$, $Ct(n_c)$, $Wd(n_c)$, or $Ty(n_c)$). Thus in the process of extracting features for \textit{CC}, for any two \textit{Cn} names $n^{(1)}_{a,b} \in \mathcal{G}^{(1)}$ and $n^{(2)}_{c,d} \in \mathcal{G}^{(2)}$, we can get their transformed forms $t(n^{(1)}_{a,b})$ and $t(n^{(2)}_{c,d})$. Since $t(n^{(1)}_{a,b})$ and $t(n^{(2)}_{c,d})$ are used to extract the similarity features of the Chinese letters in $n^{(1)}_{a,b}$ and $n^{(2)}_{c,d}$, all the English letters in $t(n^{(1)}_{a,b})$ and $t(n^{(2)}_{c,d})$ are just set to their lower cases. Then besides the 18 features that are available for \textit{EE}, we extract some new features for \textit{CC} as follows:
\begin{enumerate}[1.]
	\item \textit{The similarity between the transformed Cn names}: similar to extract the account name similarity for \textit{EE}, we use Eq. (\ref{eq:ld_sim}) to compute $sl(t(n^{(1)}_{a,b}),t(n^{(2)}_{c,d}))$, which is the similarity between $t(n^{(1)}_{a,b})$ and $t(n^{(2)}_{c,d})$. 
	\item \textit{The Proportion of the longest common substring of two transformed Cn names}: by using the Eq. (\ref{eq:pls_sim}), we can compute the proportion of the \textit{longest common substring} for $t(n^{(1)}_{a,b})$ and $t(n^{(2)}_{c,d})$. Similar to extract the proportion of the longest common substring for \textit{EE}, considering whether to eliminate the influence of word-splitting symbols in the names or not, we can extract two features here.
	\item \textit{The similarity of the abbreviations of two transformed Cn names}: by using the Eq. (\ref{eq:lcq_sim}), we can compute the similarity of the abbreviations $sa(t(n^{(1)}_{a,b}),t(n^{(2)}_{c,d}))$ from  $t(n^{(1)}_{a,b})$ and $t(n^{(2)}_{c,d})$.
	\item \textit{The similarity of the non-special letters of two transformed Cn names:} 
	for a given account name $n_a$, similar to $ns(n_a)$ which is applied to extract the non-special letters for the names matched by \textit{EE} or \textit{CE}'s, let $ns(t(n_a))$ denote the string formed by all of $t(n_a)$'s English letters and Chinese letters according to their orders in $n_a$. And we can firstly extract the strings of English and Chinese letters $ns(t(n^{(1)}_{a,b}))$ and $ns(t(n^{(2)}_{c,d}))$ from $t(n^{(1)}_{a,b})$ and $t(n^{(2)}_{c,d})$ respectively, then by using the same ways of extracting the four features of \textit{The similarity of the non-special letters} for \textit{EE}, we can extract four features for \textit{CC} from $ns(t(n^{(1)}_{a,b}))$ and $ns(t(n^{(2)}_{c,d}))$.
\end{enumerate}
So we can extract $8$ new features for \textit{CC} when given a name transform function $t(n_c)$. And since $t(n_c)$ can be $Ts(n_c)$, $Hy(n_c)$, $Ct(n_c)$, $Wd(n_c)$, or $Ty(n_c)$, we can get $8\times5$ new features. Considering the $18$ features for \textit{EE} that can be used in \textit{CC}, so in total there are $40+18$ features for \textit{CC}.

\subsection{The framework of \textit{\underline{M}ulti-View \underline{C}ross-Network \underline{U}ser \underline{A}lignment (MCUA)}}
In real world, a user account $u^{(a)}_i$ in a given network $\mathcal{G}^{(a)}$ can have a name set $N^{(a)}_i$, which may contain one or more names (e.g., $N^{(a)}_i=\{ n^{(k)}_{i,1}, n^{(k)}_{i,2},\dots\}$). For example, a twitter user account can have a full name like \textit{Jack\_Wu} and a screen name like \textit{JackWu123}. So if each account in network $\mathcal{G}^{(1)}$ has two names, and each account in network $\mathcal{G}^{(2)}$ has two names, we can conduct $2 \times 2$ times of name matchings when trying to align any two accounts between $\mathcal{G}^{(1)}$ and $\mathcal{G}^{(2)}$. Besides, as we discussed before, there exist three kinds of Chinese user account name matchings (\textit{EE}, \textit{CE}, and \textit{CC}). So in this subsection, we design a classifier-level fusion based multi-view framework \textit{MCUA} which can integrate all the results returned by different name matching models in each time of name matchings, and then generate a unified result to predict whether two given Chinese user accounts belong to the same user. 

\begin{figure*}
	\centering
	\includegraphics[width=14.0cm,height=7.4cm]{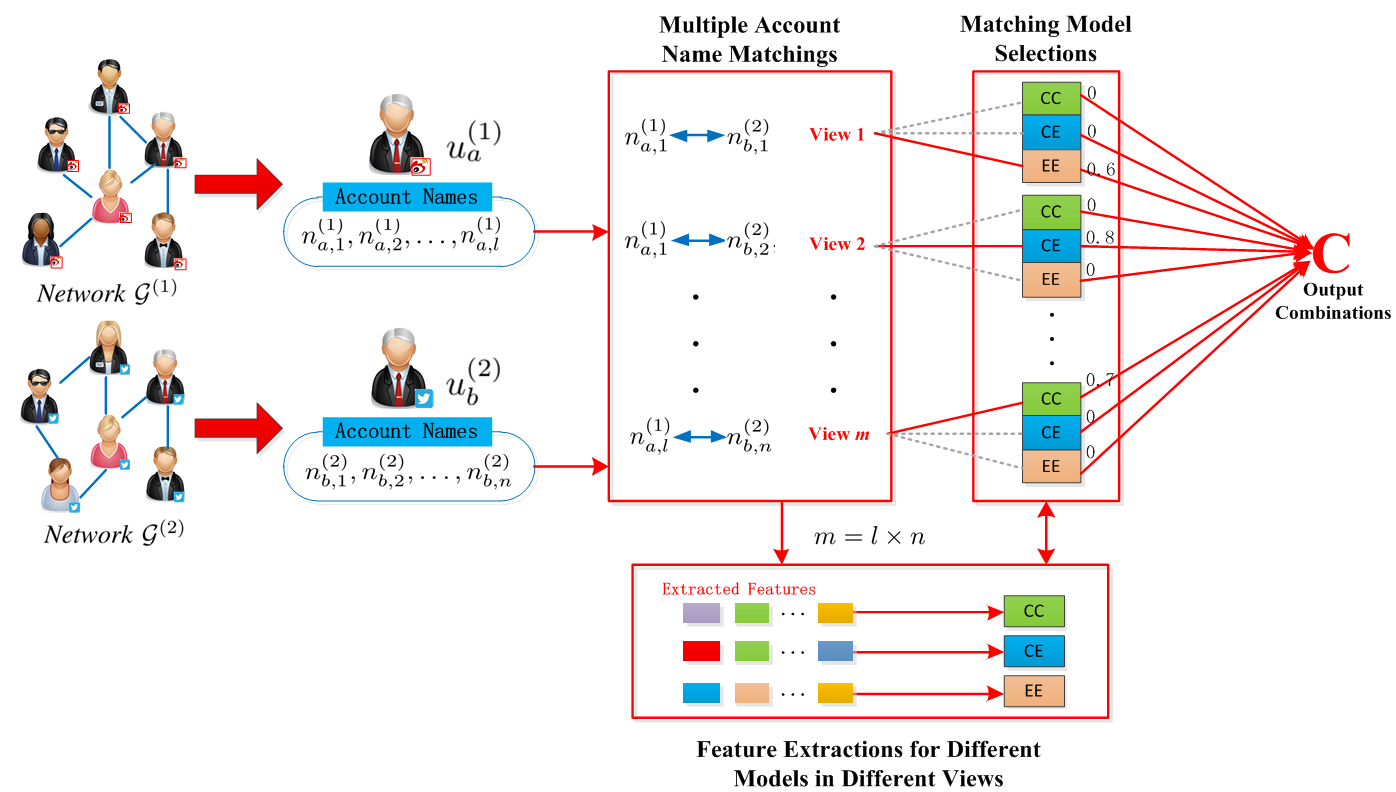}
	\caption{The framework of \textit{\underline{M}ulti-View \underline{C}ross-Network \underline{U}ser \underline{A}lignment (MCUA)}.}
	\label{fig:multi-view}
\end{figure*}

Figure \ref{fig:multi-view} illustrates the framework of \textit{MCUA}. According to it, if each Chinese user account in network $\mathcal{G}^{(1)}$ has $l$ names, and each Chinese user account in network $\mathcal{G}^{(2)}$ has $n$ names. Then when trying to align the $a$th user account $u^{(1)}_a$ in $\mathcal{G}^{(1)}$ with the $b$th user account $u^{(2)}_b$ in $\mathcal{G}^{(2)}$, there will be $m=l \times n$ pairs of names for matching. Let each pair of names be matched in one view, and thus each view should select one type of Chinese name matchings from \textit{EE}, \textit{CE}, and \textit{CC} to match the names. As we see in Figure \ref{fig:multi-view}, if both $n^{(1)}_{a,1}$ and $n^{(2)}_{b,1}$ are \textit{En} names and should be matched in \textit{View 1}, then we connect \textit{View 1} with \textit{EE} by the red solid line, which means $n^{(1)}_{a,1}$ and $n^{(2)}_{b,1}$ will be matched by \textit{EE} in \textit{View 1}. And we use gray dotted lines to connect \textit{View 1} with \textit{CE} and \textit{CC}, in this way to denote that $n^{(1)}_{a,1}$ and $n^{(2)}_{b,1}$ will not be matched by \textit{CE} and \textit{CC} in \textit{View 1}. The relationships of the other views with the three types of Chinese name matchings are illustrated in the same way. And at the same time, many similarity features of the account name pairs should be extracted for the models of \textit{EE}, \textit{CE}, and \textit{CC} to conduct name matchings. Here we use the rectangles with different colors to denote different types of extracted features. By using these features, the selected name matching model in each view can output the predicted probability of $u^{(1)}_a$ and $u^{(2)}_b$ belonging to the same user. For those name matching models that are not selected in each view, their outputs are set as 0. Since for different name matching models selected by different views, the predicted probabilities of $u^{(1)}_a$ and $u^{(2)}_b$ are owned by the same user can be different, \textit{MCUA} considers the outputs of all the name matching models by inputting them into a given classifier $C$, and using it to predict whether $u^{(1)}_a$ and $u^{(2)}_b$ belong to the same user (i.e., using it to predict the value of $A_{a,b}$).

The training of \textit{MCUA} can be divided into two steps: The first setp is to train the models for \textit{CC}, \textit{CE}, and \textit{EE}, and the second step is to train the classifier $C$. The training process of the models of \textit{CC}, \textit{CE}, and \textit{EE} is shown in Algorithm \ref{alg:train_1}, the main idea of which is to create the training sets for \textit{CC}, \textit{CE}, and \textit{EE} from the given set of the labeled alignment relationships between user accounts, and use them to train the models of \textit{CC}, \textit{CE}, and \textit{EE} separately. While the training process of classifier $C$ is shown in Algorithm \ref{alg:train_2}, whose main idea is to create the training set for $C$ by the outputs of the trained models of \textit{CC}, \textit{CE}, and \textit{EE} over the labeled user account alignment relationships, and use it to train $C$. Where $u^{(1)}_i$ ($u^{(2)}_j$) represents the $i$th ($j$th) user account in $\mathcal{G}^{(1)}$ ($\mathcal{G}^{(2)}$), and $n^{(1)}_{i,y}$ ($n^{(2)}_{j,z}$) represents the $y$th ($z$th) account name of $u^{(1)}_i$ ($u^{(2)}_j$).
\begin{algorithm}[htb]
	\caption{The Training Process of the Models of \textit{CC}, \textit{CE}, and \textit{EE}}
	\label{alg:train_1}
	\begin{algorithmic}[1]
		\Require
		$\mathcal{G}^{(1)}$, $\mathcal{G}^{(2)}$: two networks;
		$\mathcal{A}_r \in \mathcal{A}$: a set of labeled alignment relationships between the Chinese user accounts in $\mathcal{G}^{(1)}$ and $\mathcal{G}^{(2)}$;
		$F_{CC}$, $F_{CE}$, $F_{EE}$: the feature lists used by the models of \textit{CC}, \textit{CE}, and \textit{EE} separately;
		\Ensure $M_{CC}$, $M_{CE}$, $M_{EE}$: the trained models of \textit{CC}, \textit{CE}, and \textit{EE} separately;\\
		
		Initialize $3$ empty training sets: $S_{CC}$, $S_{CE}$ and $S_{EE}$
		\State Initialize the parameters of $M_{CC}$, $M_{CE}$ and $M_{EE}$ randomly
		\For {each $A_{i,j}$ in $\mathcal{A}_r$}
		\State select its related user account $u^{(1)}_i \in \mathcal{G}^{(1)}$ and $u^{(2)}_j \in \mathcal{G}^{(2)}$, and find their name sets $N^{(1)}_i$ and $N^{(2)}_j$.
		\For {each name $n^{(1)}_{i,y}$ in $N^{(1)}_i$}
		\For {each name $n^{(2)}_{j,z}$ in $N^{(2)}_j$}
		\State Use $n^{(1)}_{i,y}$, $n^{(2)}_{j,z}$ to create the name pair $p_{i,y,j,z}$
		\If {both $n^{(1)}_{i,y}$ and $n^{(2)}_{j,z}$ are \textit{En} names}
		\State Extract the features in $F_{EE}$ from $p_{i,y,j,z}$, and using them to form a feature vector $v_f$
		\State Add $\{v_f, A_{i,j}\}$ to $S_{EE}$
		\ElsIf {both $n^{(1)}_{i,y}$ and $n^{(2)}_{j,z}$ are \textit{Cn} names}
		\State Extract the features in $F_{CC}$ from $p_{i,y,j,z}$, and using them to form a feature vector $v_f$
		\State Add $\{vf, A_{i,j}\}$ to $S_{CC}$
		\Else
		\State Extract the features in $F_{CE}$ from $p_{i,y,j,z}$, and using them to form a feature vector $v_f$
		\State Add $\{v_f, A_{i,j}\}$ to $S_{CE}$
		\EndIf
		\EndFor
		\EndFor
		\EndFor\\		
		Train the name matching model $M_{CC}$ on $S_{CC}$, $M_{CE}$ on $S_{CE}$ and $M_{EE}$ on $S_{EE}$ until convergence;
	\end{algorithmic}
\end{algorithm}

\begin{algorithm}[htb]
	\caption{The Training Process of the Classifier $C$.}
	\label{alg:train_2}
	\begin{algorithmic}[1]
		\Require
		$\mathcal{G}^{(1)}$, $\mathcal{G}^{(2)}$: two networks;
		$\mathcal{A}_r \in \mathcal{A}$: a set of labeled alignment relationships between the Chinese user accounts in $\mathcal{G}^{(1)}$ and $\mathcal{G}^{(2)}$;
		$M_{CC}$, $M_{CE}$, $M_{EE}$: the trained models of \textit{CC}, \textit{CE}, and \textit{EE} separately;
		$F_{CC}$, $F_{CE}$, $F_{EE}$: the feature lists used by $M_{CC}$, $M_{CE}$ and $M_{EE}$ separately;
		$l$, $n$: the numbers of account names for each user in $\mathcal{G}^{(1)}$ and $\mathcal{G}^{(2)}$ separately;
		\Ensure $C$: the trained classifier;\\
		
		Initialize an empty training set $S_{C}$ for the classifier $C$.
		\For {each $A_{i,j}$ in $\mathcal{A}_r$}
		\State select its related user account $u^{(1)}_i \in \mathcal{G}^{(1)}$ and $u^{(2)}_j \in \mathcal{G}^{(2)}$, and find their name sets $N^{(1)}_i$ and $N^{(2)}_j$.
		\State Create a vector $v_c$ whose length is $3*l*n$, and set all the values in it as $0$
		\For {each name $n^{(1)}_{i,y}$ in $N^{(1)}_i$}
		\For {each name $n^{(2)}_{j,z}$ in $N^{(2)}_j$}
		\State Use $n^{(1)}_{i,y}$, $n^{(2)}_{j,z}$ to create the name pair $p_{i,y,j,z}$, and set $b=3(y-1)*n+3(z-1)$
		\If {both $n^{(1)}_{i,y}$ and $n^{(2)}_{j,z}$ are \textit{Cn} names}
		\State Extract the features in $F_{CC}$ from $p_{i,y,j,z}$, and input them into $M_{CC}$
		\State Set $v_c[b]$ as the output value of $M_{CC}$
		\ElsIf {both $n^{(1)}_{i,y}$ and $n^{(2)}_{j,z}$ are \textit{En} names}
		\State Extract the features in $F_{EE}$ from $p_{i,y,j,z}$, and input them into $M_{EE}$
		\State Set $v_c[b+2]$ as the output value of $M_{EE}$
		\Else
		\State Extract the features in $F_{CE}$ from $p_{i,y,j,z}$, and input them into $M_{CE}$
		\State Set $v_c[b+1]$ as the output value of $M_{CE}$
		\EndIf
		\EndFor
		\EndFor\\
		Add $\{v_c, A_{i,j}\}$ to $S_{C}$
		\EndFor\\
		Train the classifier $C$ on $S_{C}$ until convergence;
	\end{algorithmic}
\end{algorithm}

\section{Experiment}
\label{sec:exp}
In this section, we first introduce the data sets for the experiments, and then present experimental results as well as empirical analysis.

\subsection{Data Preparation}
We crawl our experimental datasets from two HINs. One is Sina Weibo, which is a Chinese microblogging (weibo) website mainly used by Chinese users. It is one of the most popular social media sites in China, in use by over 30\% of Internet users. And about 100 million messages are posted each day on Sina Weibo. Sina executives invited and persuaded many Chinese celebrities to join the platform.
The other is Twitter, an online news and social networking service where users post and interact with messages. Twitter Inc. is based in San Francisco, California, United States, and has more than 25 offices around the world. It is one of the most largest online social networks in the world, and used by users from different countries.

\subsubsection{The data samples from Sina Weibo}
Not all the accounts in Sina Weibo are good samples for this study, because there exist a lot of user accounts owned by spammers or paid posters \cite{Wang2015Detecting}, and most of the time the names of these accounts are generated by programs or generated casually by their owners without considering any meaning or naming habit. Therefore, the features extracted from these account names can be very different from the features of normal Chinese account names, which means the accounts of spammers are bad samples to our study. However, we observe that people should offer the information of their real-life identities to Sina Weibo if they want to become the verified users. But spammers or paid posters are likely to have a large number of accounts and need to act as many different roles. Besides, their accounts are often banned by the social network administrators, thus they are not likely to provide their real-life information and become the verified users. So we collect the accounts of verified users in Sina Weibo to avoid picking up bad samples owned by spammers or paid poster. And since the Sina Weibo username which is used to login by one user is invisible to the other people, here we only consider the screen name (for Sina users, it is also referred as the nickname) of each account in Sina Weibo.
\subsubsection{The data samples from Twitter}
To crawl good account samples from Twitter for our study, we should distinguish the Chinese Twitter accounts firstly. As Chinese users' profiles often contain Chinese letters, we need to pay attention to the accounts whose profiles contain Chinese letters. However, we realize that Japanese users may also employ Chinese words in their profiles, and thus the user whose profile contains Japanese letters should be filtered. Here, we crawl two kinds of account names from Twitter, one is screen name, the other is full name (for twitter users, it is also referred as the nickname). Screen name varies from account to account, it should be only composed from English letters, numbers and ``\_''. In many cases, screen name can play the same role as user ID. For example, a man can tweet message that begin with a symbol @, followed by a user screen name \cite{Chandra2012Estimating}, then the user with this screen name will be informed of this message. Full name, on the other hand, can be composed of any letters including Chinese letters, and some different accounts can share the same full name. Unlike Sina Weibo, since a lot of user accounts used by real Chinese users are not verified on Twitter, we can only filter the bad account samples owned by spammers or paid posters via distinguishing some of their typical features listed in \cite{Wang2015Detecting,Kolari2006Detecting}. 

Our ground truth data samples of the account alignment relationships between Sina Weibo and Twitter are acquired in the following ways:

\begin{enumerate}[1)]
	\item Some websites (such as about.me, Blogcatalog) may be used by some users to list their blogs, microblogs, and other social network homepages in order to attract more people to know them. By exploring the homepages of Twitter and Sina Weibo listed on these websites by each Chinese user, the alignment relationship of accounts owned by each user will be known.
	\item Some Chinese Twitter users prefer to list their Sina Weibo accounts on their Twitter Profiles, thus the alignment relationships of accounts can be acquired.
	\item Some Chinese users of Sina Weibo (or Twitter) may list their accounts of Twitter (or Sina Weibo) in their generated contents, such as tweets, comments or replies. Thus the alignment relationships of their accounts can be extracted according to these contents.
\end{enumerate}
And thus we capture 1709 positive alignment relationships of accounts between Sina Weibo and Twitter. All of these 1709 relationships belong to different users and will be used to form the experimental positive sample set. For negative samples, we construct each of them by randomly creating a user account alignment relationship $A_{i,j}$, which connect two accounts $u^{1}_i \in \mathcal{G}^{(1)}$ and $u^{2}_j \in \mathcal{G}^{(2)}$. Where $u^{1}_i$ is connected by one positive sample, and $u^{2}_j$ is connected by a different positive sample to guarantee that they do not belong to the same user. In this way, we generate up to $1709*1708$ negative user account alignment relationships. Moreover, we notice that in real-world user account alignment problem, the data samples are usually imbalance, where the negative samples can be more than the positive samples. So in our experiments, we randomly sample the negative samples from these generated relationships according to the predefined data imbalance rate ($R_{NP}$, \mbox{$R_{NP}=\frac{\#negative\_pairs}{\#positive\_pairs}$}), and use the sampled negative samples to form the experimental negative sample set. And in each group of our experiments, we assign $R_{NP}$ with different values, so that to study the performances of our method under different data imbalance rate. Finally, we divide all of our experimental samples into two parts with 5 folds cross validations: 1 fold as the training set, which is used to train the user account alignment models; and the other 4 folds as the test set, which is used to test the performances of the trained models.

\subsection{The performances comparisons}
In this subsection, we conduct a group of experiments to evaluate the performances of our \textit{MCUA} on aligning Chinese user accounts by using the account name information. By setting the data imbalance rate with different values (i.e., $R_{NP}=\{1,2,5,10,20,40\}$), we generate different data sets to conduct the experiments. And we select six methods which use name information to connect user accounts as the base-line methods. So in total, there are seven methods to be compared. The compared methods are summarized as follows:
\begin{itemize}
\item \textit{\underline{M}ulti-View \underline{C}ross-Network \underline{U}ser \underline{A}lignment (MCUA)}: our proposed multi-view approach. We set its learning model of \textit{CC} as the $l$2-Regularized $l$2-Loss SVM, set its learning model of \textit{CE} as the Random Forest, and set its learning models of \textit{EE} as well as the classifier \textit{C} as the $l$1-Regularized Logistic Regression.
\item \textit{OM-LR}: a state-of-art user account matching method which is based on $l$1-Regularized Logistic Regression and can perform very well on matching the Chinese user accounts \cite{Zhang2016Social}. It utilizes a new feature which can be used to deal with three common cases of Chinese nickname matching effectively. And for fair comparisons, we assume that the user accounts are only aligned by the extracted account name information without using any other information.
\item \textit{Content-based method}: the state-of-art username matching method used in \cite{zhang2015cosnet}. The main idea of it is to use TF-IDF to covert the name(s) of each user account into a weighted vector, and then use these vectors to compute the account similarities. These computed similarities will to be used to judge whether two accounts belong to the same user.
\item \textit{Simple-EE}: this method is not a multi-view approach. It just directly trains a learning model from a group of given features for user account alignment as many traditional approaches do \cite{Zafarani2013Connecting,liu2014hydra,lu2014identifying}. Where the given features are the available features that can be used in \textit{EE} according to this study, and are extracted from all of the nicknames.
\item \textit{Simple-CE}: this method is similar to \textit{Simple-EE}, and the features used by it are the available features that can be used in \textit{CE} according to this study, and are extracted from all of the nicknames.
\item \textit{Simple-CC}: this method is similar to \textit{Simple-EE} and \textit{Simple-CE}, and the features used by it are the available features that can be used in \textit{CC} according to this study, and are extracted from all of the nicknames.
\item \textit{Simple-All}: This method extracts all the features that are studied in this paper from all the names pairs of any two given user accounts in different networks, and use these extracted features to train a learning model for user account alignment as many traditional approaches do \cite{Zafarani2013Connecting,liu2014hydra,lu2014identifying}. Since it is not a multi-view approach and doesn't consider that different models may adapt to different types of name matchings, it can be regarded as a simplified form of \textit{MCUA}.
\end{itemize}

In this group of experiments, since \textit{MCUA} and \textit{OM-LR} use the $l$1-Regularized Logistic Regression as the learning model to determine whether two given accounts belong to the same user, to make fair comparisons, we set the learning model of the other 5 compared methods as the $l$1-Regularized Logistic Regression. In order to evaluate the performances of these compared methods on aligning user accounts, we select three different metrics in terms of F1-measure (F1), Precision (Prec.), Recall (Rec.), and the results are shown in Table \ref{tab:exp1}, in which the best performances are listed in bold.
\begin{table*}[htb]
\centering
\newcommand{\tabincell}[2]{\begin{tabular}{@{}#1@{}}#2\end{tabular}}
\caption{The performances of different account alignment methods over the data sets with different $R_{NP}$ values}
\label{tab:exp1}
\resizebox{\textwidth}{50mm}{
\begin{tabular}{clcccccc}
	\toprule[1pt]
	\multirow{2}{*}{\tabincell{c}{The Metric}} &   \multirow{2}{*}{Method}    &       \multicolumn{6}{c}{The data imbalance rate $R_{NP}$} \\
	\cline{3-8}
	& & 1 & 2 & 5& 10 & 20 & 40 \\
	\toprule
	\multirow{7}{*}{Prec.}   & \textit{Simple-EE} & 0.950718 & 0.971527 & 0.981520 & 0.977548 & 0.965207 & 0.960643 \\
	& \textit{Simple-CE} & 0.951621 & 0.965025 & 0.976748 & 0.971143 & 0.960239 & 0.954481 \\
	& \textit{Simple-CC} & 0.959671 & 0.979295 & 0.977346 & 0.974444 & 0.963651 & 0.965005 \\
	& \textit{Simple-All} &0.977916&0.976502&0.972484&0.966372&0.955133&0.958258\\
	& \textit{Content-based method} & \textbf{0.979750} & 0.983517 & 0.978133 & 0.966314 & 0.953062 & 0.952771 \\
	& \textit{OM-LR} & 0.963095 & \textbf{0.996110} & \textbf{0.993634} & \textbf{0.983232} & \textbf{0.971000} & \textbf{0.971095} \\
	& \textit{MCUA} & 0.963349 &0.973401&0.968704&0.965351&0.950115&0.953486 \\
	\toprule
	\multirow{7}{*}{F1}   & \textit{Simple-EE} & 0.838082 & 0.828837 & 0.810909 & 0.790853 & 0.772626 & 0.766060 \\
	& \textit{Simple-CE} & 0.847076 & 0.83432 & 0.812306 & 0.792648 & 0.772113 & 0.767309 \\
	& \textit{Simple-CC} & 0.841381 & 0.837645 & 0.818637 & 0.801404 & 0.784024 & 0.779779 \\
	& \textit{Simple-All} &0.902996&0.895568&0.875860&0.856749&0.834926&0.833440\\
	& \textit{Content-based method} & 0.882155 & 0.877601 & 0.862590 & 0.838979 & 0.815519 & 0.814415 \\
	& \textit{OM-LR} & 0.712044 & 0.623438 & 0.585863 & 0.584038 & 0.581866 & 0.581421 \\
	& \textit{MCUA} & \textbf{0.912799} & \textbf{0.910794} & \textbf{0.891560} & \textbf{0.882685} & \textbf{0.865396} & \textbf{0.863976} \\
	\toprule
	\multirow{7}{*}{Rec.}   & \textit{Simple-EE} & 0.749561 & 0.722765 & 0.690876 & 0.664081 & 0.644152 & 0.637286 \\
	& \textit{Simple-CE} & 0.763512 & 0.734944 & 0.695304 & 0.669617 & 0.645702 & 0.641715 \\
	& \textit{Simple-CC} & 0.749339 & 0.731845 & 0.704384 & 0.680688 & 0.66098 & 0.654556 \\
	& \textit{Simple-All} &0.838798&0.827062&0.796723&0.769486&0.741803&0.737817\\
	& \textit{Content-based method} & 0.802260 & 0.792295 & 0.771479 & 0.741364 & 0.712796 & 0.711687 \\
	& \textit{OM-LR} & 0.565319 & 0.453718 & 0.415410 & 0.415410 & 0.415410 & 0.414967 \\
	& \textit{MCUA} & \textbf{0.867365} & \textbf{0.855848} & \textbf{0.825956} & \textbf{0.813110} & \textbf{0.794729} & \textbf{0.790301} \\		
	\bottomrule[1pt]
\end{tabular}}%
\label{tab:addlabel}%
\end{table*}%

According to the results in Table \ref{tab:exp1}, we can conclude that:
\begin{itemize}
\item All the compared methods have very similar precision values on the same experimental data sets. It means that to the performance comparisons, the differences on the F1 values and Recall values are more decisive.
\item By properly extracting different features for different types of Chinese name matchings, and constructing a multi-view framework to conside ring the information of all the name pairs generated from any two user accounts' name lists, our \textit{MCUA} can significantly outperform the state-of-art \textit{Content-based method} and \textit{OM-LR} method on the F1 values and Recall values.
\item According to the F1 values and Recall values, \textit{MCUA} can significantly outperform \textit{Simple-EE}, \textit{Simple-CE}, \textit{Simple-CC} and \textit{Simple-All}. It means that building a multi-view framework, which extracts different feature groups for different types of name matchings, and can assign different proper models to deal with different feature groups, seems more reasonable to the name-based Chinese account matchings.
\end{itemize}

\subsection{Choosing the best learning model}
\label{sec:exp_model_choice}
To choose the best learning model for \textit{CC}, \textit{CE}, \textit{EE} and the classifier \textit{C} of our \textit{MCUA} framework respectively, we perform classification tasks using a range of learning techniques. By setting the data imbalance rate with different values (i.e., $R_{NP}=\{1,2,5,10,20,40\}$), we generate different imbalanced data sets for \textit{MCUA} to conduct the experiments. Since in one experiment, a learning model $m$ may be a good choice according to the precision values but may not be good enough according to the recall values, while the F1-measure considers both the precision and the recall when computing the values, we set the F1-measure as the only metric when selecting the best learning model.

In the experiments, from each account alignment relationship in the training (test) set of \textit{MCUA}, we can extract two name pairs, one is formed by a Sina Weibo account's screen name and a twitter account's screen name, while the other is formed by a Sina Weibo account's screen name and a twitter account's full name. And for each name pair, if it is formed by two \textit{Cn} names, we will add it to the training (test) set for studying the model of \textit{CC}; and if it is formed by two \textit{En} names, we will add it to the training (test) set for studying the model of \textit{EE}; and if it is formed by an \textit{En} name and a \textit{Cn} name, we will add it to the training (test) set for studying the model of \textit{CE}.

Here we first study the best learning models for \textit{CC}, \textit{CE} and \textit{EE} over different imbalance data sets. The results are shown in Table \ref{tab:exp21}, in which the best performances are listed in bold. And for a given type of Chinese name matchings, the average rank of each learning model is averaged over its performance ranks on different imbalanced data sets (e.g., for \textit{CC}, Navie Bayes' list of performance ranks on different imbalanced data sets is $\{5,5,6,6,7,7\}$, so its average rank over these $6$ imbalanced data sets is $(5+5+6+6+7+7)/6$). From the results, we can conclude that:
\begin{enumerate}
\item For \textit{CC}, the studied Random Forest, SVM models, and Logistic Regression models show very similar performances, and can outperform the Navie Bayes and the CART models in most cases. Besides, the $l$2-Regularized $l$2-Loss SVM has the best average rank. So we choose $l$2-Regularized $l$2-Loss SVM as the best learning model for \textit{CC}.
\item For \textit{CE}, according to the average rank, the $l$2-Regularized Logistic Regression and the Random Forest are better than the other methods. However, although these two models have very similar performances when $R_{NP} \le 2$, with the increase of $R_{NP}$, the performance of $l$2-Regularized Logistic Regression declines more significantly than the Random Forest. So we choose the Random Forest as the best learning model for \textit{CE}.
\item For \textit{EE}, the $l$1-Regularized Logistic Regression outperforms other methods in most circumstances and has the best average rank. So we choose the $l$1-Regularized Logistic Regression as the best learning model for \textit{EE}.
\end{enumerate}

\subsection{Feature importance analysis}
In Section \ref{sec:MUCA}, we studies $18+40+64$ account name features for Chinese user account alignment. However, in real world applications, using all of these features to align a large number of user accounts will cost a lot. So in this subsection, we study the importance of different features in learning the models of each type of Chinese account name matchings (i.e., \textit{CC}, \textit{CE} and \textit{EE}), and select the most valuable features. In other words, for each type of Chinese account name matchings, we try to find features that contribute the most to its name matching task. Here we use the selected best learning models for \textit{CC}, \textit{CE} and \textit{EE} in subsection \ref{sec:exp_model_choice} as the studied models. And although we have used different imbalance rates to generate different data sets, according to our experimental results, we find the influence of data imbalance rate on the feature importance is negligible. Thus we use the data set with $R_{NP}=40$ to conduct the experiments of feature importance analysis.

\begin{figure*}
	\centering
	\subfigure[\textit{CC}, Precision]{
		\begin{minipage}[b]{0.3\linewidth}
			\includegraphics[width=5.0cm,height=3.5cm]{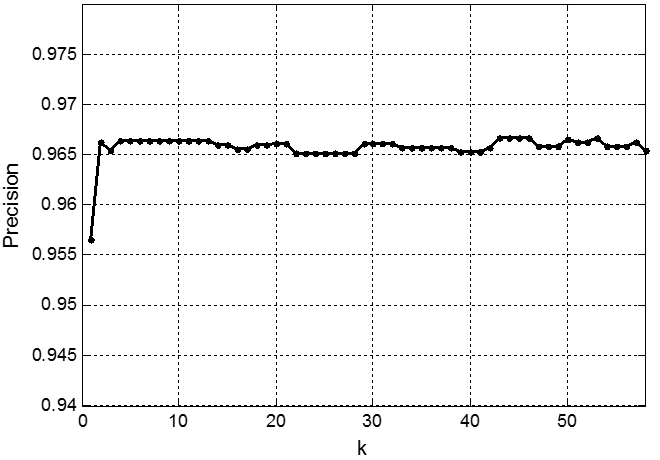}
		\end{minipage}
	}
	\subfigure[\textit{CC}, F1]{
		\begin{minipage}[b]{0.30\linewidth}
			\includegraphics[width=5.0cm,height=3.5cm]{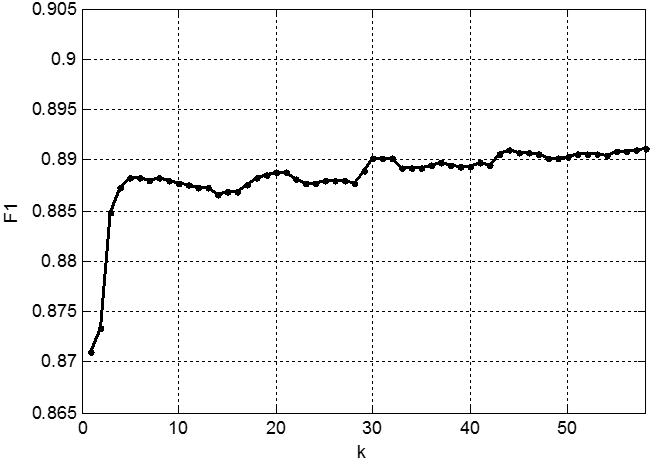}
		\end{minipage}
	}
	\subfigure[\textit{CC}, Recall]{
		\begin{minipage}[b]{0.30\linewidth}
			\includegraphics[width=5.0cm,height=3.5cm]{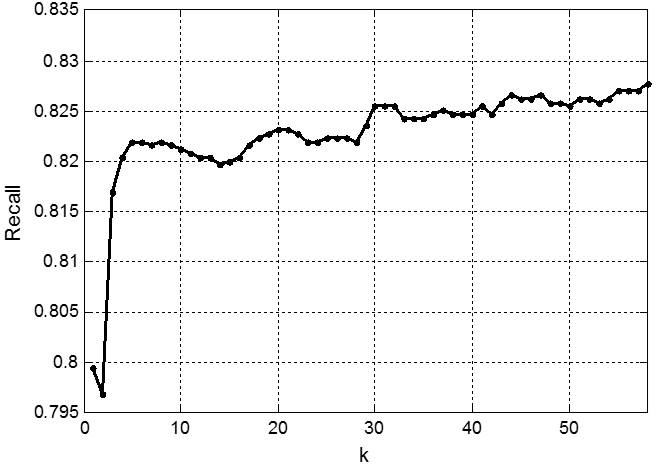}
		\end{minipage}
	}
	\subfigure[\textit{CE}, Precision]{
		\begin{minipage}[b]{0.3\linewidth}
			\includegraphics[width=5.0cm,height=3.5cm]{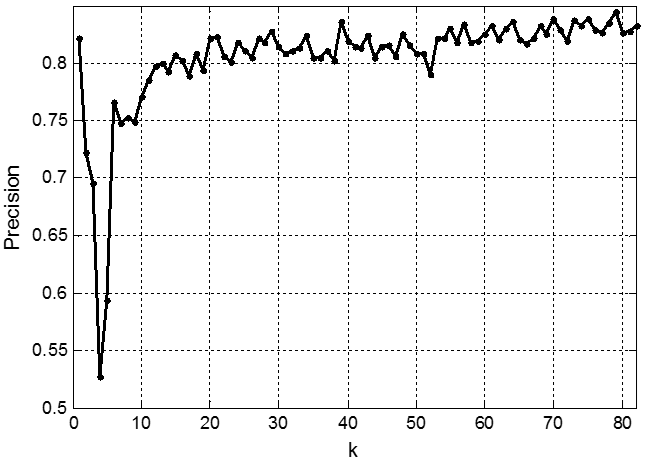}
		\end{minipage}
	}
	\subfigure[\textit{CE}, F1]{
		\begin{minipage}[b]{0.30\linewidth}
			\includegraphics[width=5.0cm,height=3.5cm]{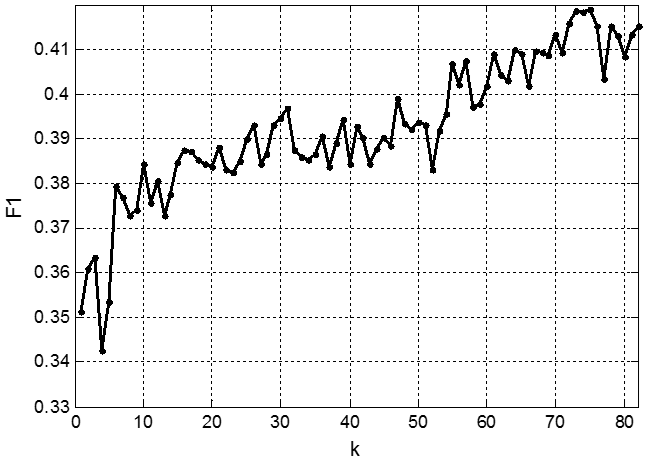}
		\end{minipage}
	}
	\subfigure[\textit{CE}, Recall]{
		\begin{minipage}[b]{0.30\linewidth}
			\includegraphics[width=5.0cm,height=3.5cm]{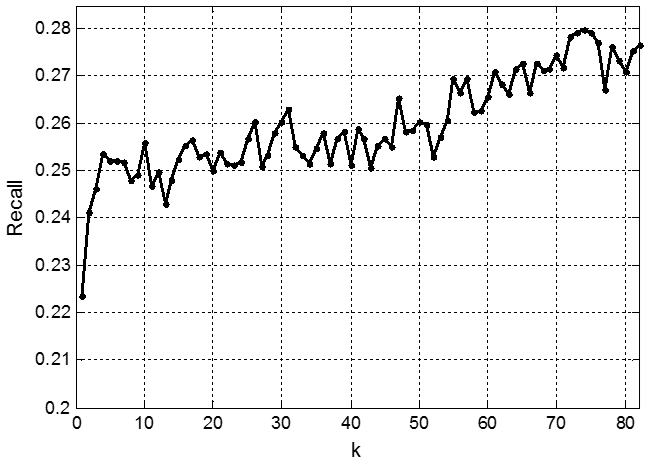}
		\end{minipage}
	}
	\subfigure[\textit{EE}, Precision]{
		\begin{minipage}[b]{0.3\linewidth}
			\includegraphics[width=5.0cm,height=3.5cm]{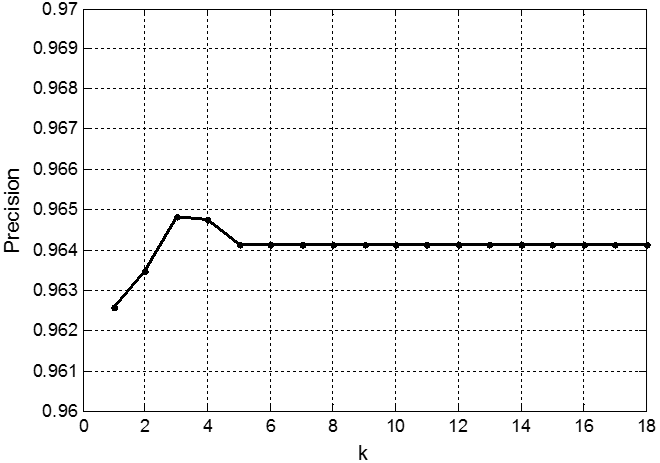}
		\end{minipage}
	}
	\subfigure[\textit{EE}, F1]{
		\begin{minipage}[b]{0.30\linewidth}
			\includegraphics[width=5.0cm,height=3.5cm]{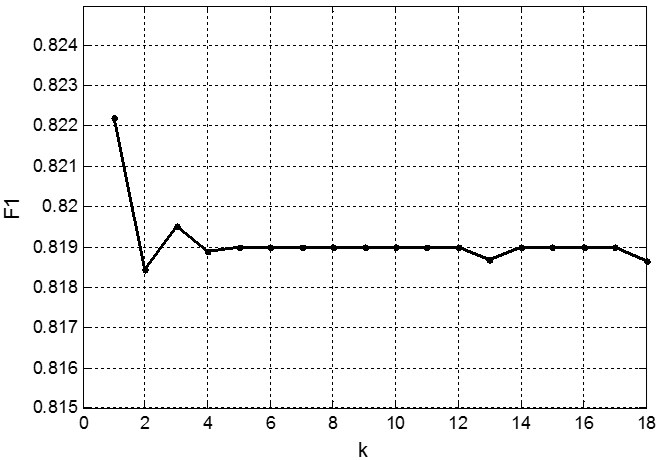}
		\end{minipage}
	}
	\subfigure[\textit{EE}, Recall]{
		\begin{minipage}[b]{0.30\linewidth}
			\includegraphics[width=5.0cm,height=3.5cm]{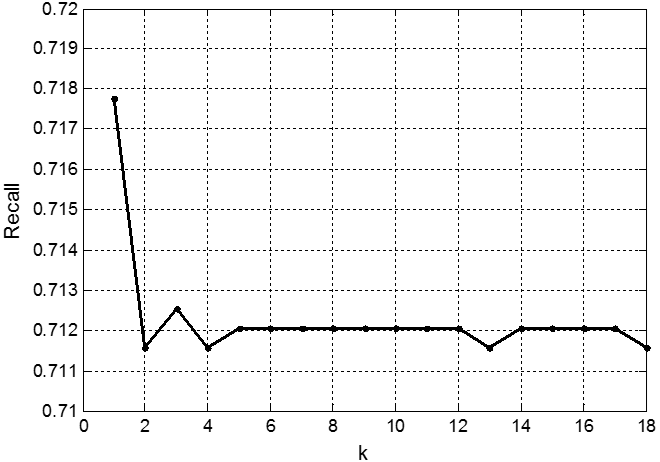}
		\end{minipage}
	}
	\caption{The performances of the learning models of \textit{CC}, \textit{CE} and \textit{EE}, when using their top-$k$ important features with different $k$ values} \label{fig:features1}
\end{figure*}

The feature importance analysis can be performed by many different feature selection measures, such as Information Gain and Pearson Correlation. And for \textit{CC} and \textit{EE}, since their best learning models are $l$1-Regularized $l$2-Loss SVM and $l$1-Regularized Logistic Regression, we use the \textit{odds-ratios} for feature importance analysis as in the work of Zafarani Reza et al. \cite{Zafarani2013Connecting} And for \textit{CE}, since its best learning model is Random Forest, which provides a straightforward feature selection method named Mean Decrease Impurity, we use Mean Decrease Impurity to compute the feature importances. In this way, we can rank the studied features according to their importances to \textit{CC}, \textit{CE} and \textit{EE}. Figure \ref{fig:features1} shows the performances of the learning models of \textit{CC}, \textit{CE} and \textit{EE}, when using their top-$k$ important features with different $k$ values. Note that, the title of each subgraph in Figure \ref{fig:features1} is formed by the type of Chinese account name matchings and the metric.

\begin{table*}[htb]
\centering
\newcommand{\tabincell}[2]{\begin{tabular}{@{}#1@{}}#2\end{tabular}}
\caption{The classification performances ``F1-measure value (rank)'' of different learning models for \textit{CC}, \textit{CE} and \textit{EE}}
\label{tab:exp21}
\resizebox{\textwidth}{40mm}{
	\begin{tabular}{clccccccc}
		\toprule[1pt]
		\multirow{2}{*}{\tabincell{c}{The type of\\matching}} &   \multirow{2}{*}{Method}    &       \multicolumn{6}{c}{The data imbalance rate $R_{NP}$} &\multirow{2}{*}{\tabincell{c}{Average\\rank}}  \\
		\cline{3-8}
		& & 1 & 2 & 5& 10 & 20 & 40 \\
		\toprule
		\multirow{7}{*}{CC}     &Naive Bayes & 0.93867 (5) & 0.93116 (2) & 0.89078 (6) & 0.83460 (6) & 0.72709 (7) & 0.71779 (7) & 5.50000  \\
		&CART & 0.90375 (7) & 0.88499 (7) & 0.84973 (7) & 0.83074 (7) & 0.81711 (6) & 0.79754 (6) & 6.66667  \\
		&Random Forest & 0.93217 (6) & 0.93027 (5) & 0.90655 (5) & 0.90136 (4) & 0.88984 (3) & 0.89406 (1) & 4.00000  \\
		&$l$1-Regularized $l$2-Loss SVM & 0.93891 (3) & 0.93054 (4) & 0.91087 (2) & 0.90219 (2) & 0.89015 (2) & 0.88892 (3) & 2.66667  \\
		&$l$2-Regularized $l$2-Loss SVM & 0.93872 (4) & 0.93111 (3) & 0.90768 (4) & \textbf{0.90367 (1)} & \textbf{0.89325 (1)} & \textbf{0.89116 (2)} & \textbf{2.50000 } \\
		&$l$1-Regularized Logistic Regression & 0.93945 (2) & 0.93019 (6) & \textbf{0.91273 (1)} & 0.90209 (3) & 0.88948 (4) & 0.88441 (4) & 3.33333  \\
		&$l$2-Regularized Logistic Regression & \textbf{0.93979 (1)} & \textbf{0.93224 (1)} & 0.91037 (3) & 0.89975 (5) & 0.88943 (5) & 0.88311 (5) & 3.33333  \\
		\toprule
		\multirow{7}{*}{CE} &Naive Bayes & 0.56958 (4) & \textbf{0.53852 (1)} & 0.42068 (4) & 0.28929 (7) & 0.17985 (7) & 0.17043 (7) & 5.00000  \\
		&CART & \textbf{0.57621 (1)} & 0.48375 (7) & 0.36557 (7) & 0.30082 (6) & 0.25910 (6) & 0.26723 (6) & 5.50000  \\
		&Random Forest & 0.56668 (5) & 0.52351 (4) & \textbf{0.47708 (1)} & \textbf{0.43588 (1)} & \textbf{0.40702 (1)} & \textbf{0.41466 (1)} & \textbf{2.16667 } \\
		&$l$1-Regularized $l$2-Loss SVM & 0.56307 (7) & 0.51724 (6) & 0.41868 (6) & 0.35871 (5) & 0.29269 (5) & 0.29196 (5) & 5.66667  \\
		&$l$2-Regularized $l$2-Loss SVM & 0.56332 (6) & 0.51914 (5) & 0.42060 (5) & 0.36337 (4) & 0.29322 (4) & 0.29928 (4) & 4.66667  \\
		&$l$1-Regularized Logistic Regression & 0.57208 (2) & 0.52683 (3) & 0.43783 (2) & 0.39956 (2) & 0.36991 (2) & 0.37450 (2) & \textbf{2.16667 } \\
		&$l$2-Regularized Logistic Regression & 0.57124 (3) & 0.52901 (2) & 0.43780 (3) & 0.39867 (3) & 0.36716 (3) & 0.37043 (3) & 2.83333  \\
		\toprule
		\multirow{7}{*}{EE} &Naive Bayes & 0.86473 (3) & 0.85744 (5) & 0.81308 (6) & 0.73211 (7) & 0.61959 (7) & 0.59243 (7) & 5.83333  \\
		&CART & 0.82505 (7) & 0.80478 (7) & 0.76767 (7) & 0.73514 (6) & 0.71126 (6) & 0.70205 (6) & 6.50000  \\
		&Random Forest & 0.85128 (6) & 0.84620 (6) & 0.85132 (2) & 0.83047 (2) & 0.81605 (2) & 0.81677 (2) & 3.33333  \\
		&$l$1-Regularized $l$2-Loss SVM & 0.86124 (5) & 0.85972 (3) & 0.84823 (4) & 0.82986 (3) & 0.81348 (3) & 0.81167 (3) & 3.50000  \\
		&$l$2-Regularized $l$2-Loss SVM & 0.86437 (4) & 0.85870 (4) & 0.84613 (5) & 0.82891 (4) & 0.81118 (4) & 0.81048 (4) & 4.16667  \\
		&$l$1-Regularized Logistic Regression & 0.86718 (2) & \textbf{0.86148 (1)} & \textbf{0.85287 (1)} & \textbf{0.83263 (1)} & \textbf{0.81698 (1)} & \textbf{0.81866 (1)} & \textbf{1.16667 } \\
		&$l$2-Regularized Logistic Regression & \textbf{0.86792 (1)} & 0.86146 (2) & 0.85053 (3) & 0.82730 (5) & 0.80887 (5) & 0.80868 (5) & 3.50000  \\

		\bottomrule[1pt]
	\end{tabular}%
}
\label{tab:addlabel}%
\end{table*}%

And based on the selected best learning models for \textit{CC}, \textit{CE} and \textit{EE}, we further study the best learning model for the classifier \textit{C} over different imbalanced data sets, which are the generated imbalanced Chinese account alignment relationship sets for \textit{MCUA}. The results are shown in Table \ref{tab:exp22}, in which the best performances are listed in bold. And similar to Table \ref{tab:exp21}, the average rank of each learning model is averaged over its performance ranks on different imbalanced data sets. From Table \ref{tab:exp22}, we can see that the studied Random Forest, SVM models, and Logistic Regression models show very similar performances, and significantly outperform the Navie Bayes and the CART models. Besides, the $l$1-Regularized Logistic Regression has the best average rank. So we select the $l$1-Regularized Logistic Regression as the best learning model for the classifier \textit{C}.
\begin{table*}[htb]
\centering
\newcommand{\tabincell}[2]{\begin{tabular}{@{}#1@{}}#2\end{tabular}}
\caption{The classification performances ``F1-measure value (rank)'' of different learning models for the classifier \textit{C}}
\label{tab:exp22}
\resizebox{\textwidth}{16mm}{
	\begin{tabular}{lccccccc}
		\toprule[1pt]
		\multirow{2}{*}{Method}    &       \multicolumn{6}{c}{The data imbalance rate $R_{NP}$} &\multirow{2}{*}{\tabincell{c}{Average\\rank}}  \\
		\cline{2-7}
		& 1 & 2 & 5& 10 & 20 & 40 \\
		\toprule
		Naive Bayes & 0.88832 (6) & 0.89691 (6) & 0.85396 (7) & 0.80049 (7) & 0.76072 (7) & 0.69935 (7) & 6.66667  \\
		CART & 0.87586 (7) & 0.87978 (7) & 0.85506 (6) & 0.84668 (6) & 0.83234 (5) & 0.83620 (5) & 6.00000  \\
		Random Forest & 0.89246 (5) & 0.90014 (5) & 0.87439 (5) & 0.86687 (4) & 0.85092 (4) & 0.85908 (2) & 4.16667  \\
		$l$1-Regularized $l$2-Loss SVM & 0.91226 (3) & \textbf{0.91197 (1)} & 0.88785 (2) & 0.87852 (2) & 0.86295 (2) & 0.85874 (3) & 2.16667  \\
		$l$2-Regularized $l$2-Loss SVM & 0.91259 (2) & 0.91101 (2) & 0.88766 (3) & 0.87824 (3) & 0.86282 (3) & 0.85822 (4) & 2.83333  \\
		$l$1-Regularized Logistic Regression & \textbf{0.91280 (1)} & 0.91079 (3) & \textbf{0.89156 (1)} & \textbf{0.88268 (1)} & \textbf{0.86540 (1)} & \textbf{0.86398 (1)} & \textbf{1.33333} \\
		$l$2-Regularized Logistic Regression & 0.91126 (4) & 0.90546 (4) & 0.87690 (4) & 0.85861 (5) & 0.82562 (6) & 0.83305 (6) & 4.83333  \\
		\bottomrule[1pt]
\end{tabular}}
\label{tab:addlabel}%
\end{table*}%

From Figure \ref{fig:features1}, we can see for the learning model of \textit{CC}, in general, using more features means better performances. However, its precision value does not significantly increase when $k >4$, and the improvements of its F1 and recall values are negligible when using more than 5 most important features. And thus we can conclude that using the top-$5$ important features is enough for the learning model of \textit{CC} to achieve relatively good performance, these features are listed according to their importances as follows:

\begin{enumerate}[1.]
	\item The similarity of the non-special letter distribution which is directly computed from the cosine similarity of the two given names.
	\item The percentage of the common non-special letters in all the non-special letters of the two given names, where all of the upper case English letters have been transformed to the lower case letters.
	\item The proportion of the \textit{longest common substring} of two transformed Cn names, where the Chinese letters are transformed to their \textit{Hanyu Pinyin} forms.
	\item The proportion of the \textit{longest common substring} of two transformed Cn names, where the Chinese letters are transformed to their Cantonese forms.
	\item The similarity of the non-special letters of two transformed Cn names, where all of the traditional Chinese letters are transformed to the simplified Chinese letters, and the similarity is computed by the \textit{cosine similarity} method.
\end{enumerate}

For the learning model of \textit{CE} (see Figure \ref{fig:features1}), when $k \le 3$, its F1 and recall values are very small, while when $k>3$, its the performances generally fluctuating increase with the $k$ value. And when using its top-$10$ features, its F1 and recall values reach their local maximums, which are not drastically worse than its largest F1 and recall values. At the same time, its precision value is also not much worse than the best precision value. Thus we select out the top-$10$ important features for \textit{CE}, which are listed as follows according to their importances:
\begin{enumerate}[1.]
	\item The proportion of the \textit{longest common substring} in the \textit{En} name and the transformed \textit{Cn} name, where all the Chinese letters in the \textit{Cn} name have been transformed to their \textit{Hanyu Pinyin} forms.
	\item The \textit{Jaccard index} of the non-special letters in the \textit{En} name and the transformed \textit{Cn} name, where all the Chinese letters have been transformed to their \textit{Hanyu Pinyin} forms.
	\item The proportion of the \textit{longest common substring} in the \textit{En} name and the transformed \textit{Cn} name, where all the spaces letters are eliminated and all the Chinese letters in the \textit{\textit{Cn}} name have been transformed to their \textit{Hanyu Pinyin} forms.
	\item The \textit{cosine similarity} of the non-special letters in the \textit{En} name and the transformed \textit{Cn} name, where all the Chinese letters have been transformed to their \textit{Cantonese} forms.
	\item The proportion of the \textit{longest common substring} of the strings formed by the non-special letters in the \textit{En} name and the transformed \textit{Cn} name, where all the Chinese letters have been transformed to their \textit{Hanyu Pinyin} forms.
	\item The \textit{cosine similarity} of the non-special letters in the \textit{En} name and the transformed \textit{Cn} name, where all the Chinese letters have been transformed to their \textit{Hanyu Pinyin} forms.
	\item The proportion of the \textit{longest common substring} for the transformed \textit{CE} name pair, where the Chinese family name in the beginning of the \textit{\textit{Cn}} name have been swaped to its end, and all the Chinese letters have been transformed to their \textit{Hanyu Pinyin} forms.
	\item The \textit{cosine similarity} of the non-special letters in the \textit{En} name and the transformed \textit{Cn} name, where the Chinese family name at the beginning of the \textit{Cn} name has been swapped to its end, and all the Chinese letters have been transformed to their \textit{Cantonese} forms.
	\item The \textit{cosine similarity} of the non-special letters in the \textit{En} name and the transformed \textit{Cn} name, where the Chinese family name at the beginning of the \textit{Cn} name has been swapped to its end, and all the Chinese letters have been transformed to their \textit{Hanyu Pinyin} forms.
	\item The \textit{cosine similarity} of the non-special letters in the \textit{En} name and the transformed \textit{Cn} name, where all the Chinese letters have been transformed to their \textit{Wade-Giles} forms. 
\end{enumerate}
Since the \textit{cosine similarity} of two given strings will not be influenced by the orders of their letters, for any two names matched by the learning model of \textit{CE}, the 4th and 6th features in the above list have the same values as the 8th and 9th features respectively. It means that after using the 4th and 6th features, the information contained in the 8th and 9th features may become valueless for the learning model of \textit{CE}. That can also be the reason why in Figure \ref{fig:features1} compared with only using the top-7 features, the performances of the learning model of \textit{CE} can not be obviously improved by using the top-8 and top-9 features. So in the above list, we should eliminate the 8th and 9th features, and only consider the rest 8 features.

For the learning model of \textit{EE} (see Figure \ref{fig:features1}), when $k=1$, although we can get the best F1 and recall values, however, the precision value is the worst one. However, when $k=3$, we can get the best precision value and the second best F1 and recall values. So we select out the top-$3$ features for \textit{EE}, which are listed as follows according to their importances:
\begin{enumerate}[1.]
	\item The \textit{Levenshtein distance} based similarity computed from the non-special letters of two given names.
	\item The proportion of the \textit{longest common substring}, which is extracted after all the word-splitting symbols have been deleted and all the upper case letters have been transformed to the lower case letters in both of the two given names.
	\item The proportion of the \textit{longest common substring} in both $ns(n^{(1)}_{a,b})$ and $ns(n^{(2)}_{c,d})$, which are the strings formed by all of name $n^{(1)}_{a,b}$'s and name $n^{(2)}_{c,d}$'s non-special letters according to their orders in $n^{(1)}_{a,b}$ and $n^{(2)}_{c,d}$ respectively, where all of the upper case letters have been transformed to the lower case letters.
\end{enumerate}

Table \ref{tab:feature_study} lists the performance comparisons of our proposed method using all the studied features (referred as \textit{MCUA}) and our method using the selected features in this subsection (referred as \textit{MCUA-S}). From it we can see that by only using the selected top-k features for each type of matchings, the performances of \textit{MCUA-S} are not drastically worse than \textit{MCUA}, which proves that our selected features are valuable and can help our method to achieve good enough performances in different circumstances of our studied problem.
\begin{table*}[htb]
	\centering
	\newcommand{\tabincell}[2]{\begin{tabular}{@{}#1@{}}#2\end{tabular}}
	\caption{The performance comparisons of our proposed method using all the studied features (referred as \textit{MCUA}) with our method using the selected features (referred as \textit{MCUA-S}) over different imbalanced data sets}
	\label{tab:feature_study}
	\begin{tabular}{clcccccc}
		\toprule[1pt]
		\multirow{2}{*}{\tabincell{c}{The Metric}} &   \multirow{2}{*}{Method}    &       \multicolumn{6}{c}{The data imbalance rate $R_{NP}$} \\
		\cline{3-8}
		& & 1 & 2 & 5& 10 & 20 & 40 \\
		\toprule
		\multirow{2}{*}{Prec.}   & \textit{MCUA} & 0.963349 &0.973401&0.968704&0.965351&0.950115&0.953486 \\
		& \textit{MCUA-S} & 0.962203 & 0.975662 & 0.969406 & 0.959552 &	0.948111 & 0.951929 \\
		\toprule
		\multirow{2}{*}{F1}   & \textit{MCUA} & 0.912799 & 0.910794 & 0.891560 & 0.882685 & 0.865396 & 0.863976  \\
		& \textit{MCUA-S} & 0.910734 & 0.903549 & 0.888249 & 0.874102 & 0.858776 & 0.856214  \\
		\toprule
		\multirow{2}{*}{Rec.}   & \textit{MCUA} & 0.867365 & 0.855848 & 0.825956 & 0.813110 & 0.794729 & 0.790301 \\
		& \textit{MCUA-S} &0.864708 & 0.841454 & 0.819754 & 0.802700 & 0.784987 & 0.778564 \\
		\bottomrule[1pt]
	\end{tabular}%
	\label{tab:addlabel}%
\end{table*}%

\section{Conclusions}
\label{sec:conclusions}
In this paper, we propose a \textit{\underline{M}ulti-View \underline{C}ross-Network \underline{U}ser \underline{A}lignment (MCUA)} method to deal with the problem of aligning Chinese user accounts based on the account name information. Although several existing works have tried to utilize the account name information to align Chinese user accounts, none of them have detailedly studied the multiple types of Chinese account name matchings as well as their related available features. So in our paper, we firstly discuss the details of different types of Chinese account name matchings. And then for each type of matchings, we study the available naming behavioral models as well as their related features. Thirdly, we design a classifier-level fusion based multi-view framework for our \textit{MCUA} method. This framework creatively integrates the models of different types of user name matchings and can consider all of the studied features. And thus in each time of aligning Chinese user accounts, \textit{MCUA} can use different models to deal with different types of Chinese account name matchings, and then generate a unified result according to the returned results of these models. To analyze the performances of our \textit{MCUA} method, we randomly collect the Chinese user information from Sina Weibo and Twitter, and then compare \textit{MCUA} with six base-line methods. The results show that \textit{MCUA} can outperform the other compared methods on aligning Chinese user accounts between these two networks. Besides, we also study the best learning models and the top-$k$ valuable features of different matchings for \textit{MCUA} over our experimental data sets. 

Although our study provides a reasonable way of utilizing the account name information to align Chinese user accounts. However, in some cases, only using the account name information is not enough for the alignment of user accounts (e.g., some users may use very similar account names). So how to properly integrate our proposed approach with other ways of using the other information (e.g., user relationships, users' posts and comments) for account alignment will be further studied in our future works. 

\section{Acknowledgments}
This research is partially supported by National Natural Science Foundation of China (Grant No. 61802424).

\bibliography{ref}%

\begin{thebibliography}{10}
\providecommand \doibase [0]{http://dx.doi.org/}%

\bibitem{Velayudhan2019Compromised}
Velayudhan SP, Somasundaram MSB. Compromised account detection in online social
  networks: A survey. {\it Concurrency and Computation: Practice and
  Experience} 2019(1).

\bibitem{Zhu2017cikm}
Zhu J, Zhang J, He L, et al. Broad Learning Based Multi-Source Collaborative
  Recommendation. In:  {\it Proceedings of the 2017 ACM on Conference on
  Information and Knowledge Management}. ACM; 2017; New York, NY, USA\string:
  1409--1418.

\bibitem{zhang2013predicting}
Zhang J, Kong X, Philip SY. Predicting social links for new users across
  aligned heterogeneous social networks. In:  {\it 2013 IEEE 13th International
  Conference on Data Mining}. IEEE Computer Society; 2013\string: 1289--1294.

\bibitem{lu2014identifying}
Lu CT, Shuai HH, Yu PS. Identifying your customers in social networks. In:
  {\it Proceedings of the 23rd ACM International Conference on Conference on
  Information and Knowledge Management}. ACM; 2014\string: 391--400.

\bibitem{zhang2015integrated}
Zhang J, Philip SY. Integrated anchor and social link predictions across social
  networks. In:  {\it Proceedings of the 24th International Conference on
  Artificial Intelligence}. AAAI Press; 2015\string: 2125--2131.

\bibitem{Zhu2017Constrained}
Zhu J, Zhang J, Wu Q, et al. Constrained Active Learning for Anchor Link
  Prediction Across Multiple Heterogeneous Social Networks. {\it Sensors}
  2017\string; 17(8)\string: 1786.

\bibitem{wang2018user}
Wang Y, Feng C, Chen L, Yin H, Guo C, Chu Y. User identity linkage across
  social networks via linked heterogeneous network embedding. {\it World Wide
  Web} 2018\string: 1--22.

\bibitem{li2019matching}
Li Y, Peng Y, Zhang Z, Yin H, Xu Q. Matching user accounts across social
  networks based on username and display name. {\it World Wide Web}
  2019\string; 22(3)\string: 1075--1097.

\bibitem{liu2019structural}
Liu L, Li X, Cheung W, Liao L. Structural Representation Learning for User
  Alignment Across Social Networks. {\it IEEE Transactions on Knowledge and
  Data Engineering} 2019.

\bibitem{wang2019online}
Wang W, Yin H, Du X, Hua W, Li Y, Nguyen QVH. Online user representation
  learning across heterogeneous social networks. In:  {\it Proceedings of the
  42nd International ACM SIGIR Conference on Research and Development in
  Information Retrieval}. ACM; 2019\string: 545--554.

\bibitem{li2018matching}
Li Y, Zhang Z, Peng Y, Yin H, Xu Q. Matching user accounts based on user
  generated content across social networks. {\it Future Generation Computer
  Systems} 2018\string; 83\string: 104--115.

\bibitem{ZhangCikm2018}
Zhang J, Chen B, Wang X, et al. MEgo2Vec: Embedding Matched Ego Networks for
  User Alignment Across Social Networks. In:  {\it Proceedings of the 27th ACM
  International Conference on Information and Knowledge Management}. ACM;
  2018\string: 327--336.

\bibitem{Ma2017Balancing}
Ma J, Qiao Y, Hu G, et al. Balancing User Profile and Social Network Structure
  for Anchor Link Inferring across Multiple Online Social Networks. {\it IEEE
  Access} 2017\string; PP(99)\string: 1-1.

\bibitem{Zhang2016Social}
Zhang Y, Wang L, Li X, Xiao C. Social Identity Link Across Incomplete Social
  Information Sources Using Anchor Link Expansion. In:  {\it PAKDD 2016}.
  Springer International Publishing; 2016\string: 395--408.

\bibitem{zhang2015cosnet}
Zhang Y, Tang J, Yang Z, Pei J, Yu PS. Cosnet: connecting heterogeneous social
  networks with local and global consistency. In:  {\it Proceedings of the 21th
  ACM SIGKDD International Conference on Knowledge Discovery and Data Mining}.
  ACM; 2015\string: 1485--1494.

\bibitem{liu2014hydra}
Liu S, Wang S, Zhu F, Zhang J, Krishnan R. Hydra: Large-scale social identity
  linkage via heterogeneous behavior modeling. In:  {\it Proceedings of the
  2014 ACM SIGMOD international conference on Management of data}. ACM;
  2014\string: 51--62.

\bibitem{Zafarani2013Connecting}
Zafarani R, Liu H. Connecting users across social media sites: a
  behavioral-modeling approach. In:  {\it ACM SIGKDD International Conference
  on Knowledge Discovery and Data Mining}. ACM; 2013\string: 41-49.

\bibitem{Jain2013}
Jain P, Kumaraguru P, Joshi A. @i seek 'fb.me':identifying users across
  multiple online social networks. In:  {\it Proceedings of the 22nd
  international conference on World Wide Web companion}. ACM; 2013\string:
  1259-1268.

\bibitem{zafarani2009connecting}
Zafarani R, Liu H. Connecting Corresponding Identities across Communities. {\it
  ICWSM} 2009\string; 9\string: 354--357.

\bibitem{zhang2015multiple}
Zhang J, Philip SY. Multiple anonymized social networks alignment. In:  {\it
  2015 IEEE International Conference on Data Mining (ICDM)}. IEEE. IEEE
  Computer Society; 2015\string: 599--608.

\bibitem{Xu2014Information}
Xu L, Jiang C, Wang J, Yuan J. Information Security in Big Data: Privacy and
  Data Mining. {\it Access IEEE} 2014\string; 2\string: 1149-1176.

\bibitem{Zhu2017CHRS}
Zhu J, Zhang J, Zhang C, et al. CHRS: Cold Start Recommendation across Multiple
  Heterogeneous Information Networks. {\it IEEE Access} 2017\string;
  PP(99)\string: 1-1.

\bibitem{You2011SocialSearch}
You GW, Hwang SW, Nie Z, Wen JR. SocialSearch: enhancing entity search with
  social network matching. In:  {\it International Conference on Extending
  Database Technology}. ACM; 2011\string: 515-519.

\bibitem{Backes2018}
Backes T. The Impact of Name-Matching and Blocking on Author Disambiguation.
  In:  {\it Proceedings of the 27th ACM International Conference on Information
  and Knowledge Management}. ACM; 2018; New York, NY, USA\string: 803--812.

\bibitem{liu2013cross}
Liu D, Wu QY. Cross-Platform User Profile Matching in Online Social Networks.
  In:  {\it Applied Mechanics and Materials}. Trans Tech Publ; 2013\string:
  1955--1958.

\bibitem{Iofciu2010Identifying}
Iofciu T, Fankhauser P, Abel F, Bischoff K. Identifying Users Across Social
  Tagging Systems. In:  {\it International Conference on Weblogs and Social
  Media}. AAAI; 2010.

\bibitem{Motoyama2009I}
Motoyama M, Varghese G. I seek you: searching and matching individuals in
  social networks. In:  {\it Eleventh International Workshop on Web Information
  and Data Management}. ACM; 2009\string: 67-75.

\bibitem{kong2013inferring}
Kong X, Zhang J, Yu PS. Inferring anchor links across multiple heterogeneous
  social networks. In:  {\it Proceedings of the 22nd ACM international
  conference on Information \& Knowledge Management}. ACM; 2013\string:
  179--188.

\bibitem{Nie2016Identifying}
Nie Y, Jia Y, Li S, Zhu X, Li A, Zhou B. Identifying users across social
  networks based on dynamic core interests. {\it Neurocomputing} 2016\string;
  210\string: 107-115.

\bibitem{Lu2018Releasing}
Lu O, Zheng Q, Liao S, Yuan H, Jia X. Releasing Correlated Trajectories:
  Towards High Utility and Optimal Differential Privacy. {\it IEEE Transactions
  on Dependable \& Secure Computing} 2018\string; PP(99)\string: 1-1.

\bibitem{Vosecky2009User}
Vosecky J, Hong D, Shen VY. User identification across multiple social
  networks. In:  {\it International Conference on Networked Digital
  Technologies}. IEEE; 2009\string: 360-365.

\bibitem{Cao2015Tensor}
Cao B, He L, Kong X, Yu PS, Hao Z, Ragin AB. Tensor-based Multi-view Feature
  Selection with Applications to Brain Diseases. In:  {\it IEEE International
  Conference on Data Mining}. IEEE Computer Society; 2015\string: 40-49.

\bibitem{Luo2015Multiview}
Luo Y, Liu T, Tao D, Xu C. Multiview matrix completion for multilabel image
  classification. {\it IEEE Transactions on Image Processing} 2015\string;
  24(8)\string: 2355-2368.

\bibitem{Zhu2016Block}
Zhu X, Li X, Zhang S. Block-Row Sparse Multiview Multilabel Learning for Image
  Classification. {\it IEEE Transactions on Cybernetics} 2016\string;
  46(2)\string: 450.

\bibitem{Liu2015Multiview}
Liu L, Yu M, Shao L. Multiview alignment hashing for efficient image search.
  {\it IEEE Transactions on Image Processing A Publication of the IEEE Signal
  Processing Society} 2015\string; 24(3)\string: 956.

\bibitem{Nie2016Parameter}
Nie F, Li J, Li X. Parameter-free auto-weighted multiple graph learning: a
  framework for multiview clustering and semi-supervised classification. In:
  {\it International Joint Conference on Artificial Intelligence}. AAAI Press;
  2016\string: 1881-1887.

\bibitem{Mcfee2011Learning}
Mcfee B, Lanckriet G. Learning Multi-modal Similarity. {\it Journal of Machine
  Learning Research} 2011\string; 12(8)\string: 491-523.

\bibitem{Xu2015Multi}
C X, D T, C X. Multi-View Intact Space Learning. {\it IEEE transactions on
  pattern analysis and machine intelligence} 2015\string; 37(12)\string: 2531.

\bibitem{Wozniak2009Some}
Wozniak M, Jackowski K. Some Remarks on Chosen Methods of Classifier Fusion
  Based on Weighted Voting. In:  {\it International Conference on Hybrid
  Artificial Intelligence Systems}. Springer-Verlag; 2009\string: 541-548.

\bibitem{Sindhwani2005A}
Sindhwani V, Niyogi P, Belkin M. A co-regularized approach to semi-supervised
  learning with multiple views. In:  {\it Proceedings of the ICML Workshop on
  Learning with Multiple Views}. ; 2005.

\bibitem{Kludas2008Information}
Kludas J, Bruno E, Marchandmaillet S. Information Fusion in Multimedia
  Information Retrieval. {\it Lecture Notes in Computer Science} 2008\string;
  48(5)\string: 147-159.

\bibitem{Wang2015Detecting}
Wang X, Zhou B, Jia Y, Li S. Detecting Internet Hidden Paid Posters Based on
  Group and Individual Characteristics. In:  {\it International Conference on
  Web Information Systems Engineering}. Springer International Publishing;
  2015\string: 109-123.

\bibitem{Chandra2012Estimating}
Chandra S, Khan L, Muhaya FB. Estimating Twitter User Location Using Social
  Interactions--A Content Based Approach. In:  {\it IEEE Third International
  Conference on Privacy, Security, Risk and Trust}. IEEE; 2012\string: 838-843.

\bibitem{Kolari2006Detecting}
Kolari P, Java A, Finin T, Oates T, Joshi A. Detecting Spam blogs: A machine
  learning approach. In:  {\it National Conference on Artificial Intelligence
  and the Eighteenth Innovative Applications of Artificial Intelligence
  Conference}. AAAI Press; 2006.

\end{thebibliography}

\end{document}